\documentclass[aip,pof,natbib]{revtex4-1}

\usepackage{graphicx}
\usepackage{dcolumn}
\usepackage{bm}
\usepackage[utf8]{inputenc}
\usepackage[T1]{fontenc}
\usepackage{mathptmx}
\usepackage{etoolbox}
\usepackage{xcolor}
\usepackage{url}
\usepackage[normalem]{ulem}
\usepackage{subfigure}
\usepackage{amsmath}
\usepackage{booktabs}
\usepackage{epstopdf, epsfig}
\usepackage{placeins}
\usepackage{fontspec}               

\makeatletter
\def\@email#1#2{%
 \endgroup
 \patchcmd{\titleblock@produce}
  {\frontmatter@RRAPformat}
  {\frontmatter@RRAPformat{\produce@RRAP{*#1\href{mailto:#2}{#2}}}\frontmatter@RRAPformat}
  {}{}
}%
\makeatother
\begin{document}

\title[Direct numerical simulations of  multi-layer Rayleigh-Taylor instability]{Direct numerical simulations of  multi-layer Rayleigh-Taylor instability under asymmetric interfacial density stratification\\}


\author{Abhijeet Guha}%
\affiliation{Department of Mechanical Engineering, Indian Institute of Technology Dhanbad, Jharkhand-826 004, India.}

\author{Prasoon Suchandra}%
\affiliation{George W. Woodruff School of Mechanical Engineering, Georgia Institute of Technology, Atlanta, GA, 30332, USA}

\author{Aditi Sengupta}
\email{aditi@iitism.ac.in}
\affiliation{Department of Mechanical Engineering, Indian Institute of Technology Dhanbad, Jharkhand-826 004, India.}

\date{\today}

\begin{abstract}
Direct numerical simulations of thermally stratified multi-layer Rayleigh-Taylor instability (RTI) are performed to investigate the influence of interfacial Atwood number ($At$) on instability growth, interfacial coupling, turbulent mixing, and spectral characteristics. A fully compressible, three-dimensional, high-fidelity Navier-Stokes solver has been used to simulate ten cases spanning a broad range of interfacial $At$ to examine the evolution of coupled instability dynamics. The configuration consists of three superposed fluid layers with unstable density stratification across two interfaces generated through temperature variations. The temporal evolution of the flow is characterized using the maximum vorticity, mixing-layer growth, density fluctuations, turbulent kinetic energy (TKE), and energy partitioning between gravitational potential energy, TKE, and dissipation. The instability evolution is shown to proceed through distinct onset, quasi-exponential growth, and nonlinear saturation stages for all configurations. Increasing interfacial $At$ leads to earlier instability onset, enhanced vorticity amplification, and accelerated development of turbulent mixing layers. The weakly stratified configurations exhibit dynamics resembling effectively single-interface RTI, whereas the strongly stratified cases develop two simultaneously active mixing layers that subsequently interact and merge into a coupled turbulent region. The wall-normal distributions of density fluctuations and TKE demonstrate progressive broadening of the mixing region and increasing redistribution of turbulent activity with increasing stratification strength. The late-stage spectral characteristics are examined using one-dimensional Fourier spectra of the vorticity field. The spectra exhibit an approximate -11/5 scaling over an intermediate range of wavenumbers, consistent with buoyancy-dominated turbulent transport.

\end{abstract}

\maketitle

\section{Introduction \label{sec1}}

Rayleigh-Taylor instability (RTI) is a canonical buoyancy-driven hydrodynamic instability that develops when a denser fluid is accelerated into a lighter fluid, or equivalently when a heavy fluid overlies a lighter one in the presence of gravity. The instability originates from the unstable alignment of the pressure and density gradients, such that $\nabla p \cdot \nabla \rho < 0$, leading to baroclinic vorticity generation and amplification of initially small perturbations at the interface \citep{kull1991theory, taylor1950instability, chandrasekhar1961hydrodynamic}. Owing to its fundamental role in buoyancy-driven turbulent mixing, RTI has been studied extensively in a broad range of applications including atmospheric and oceanic flows, astrophysical convection, inertial confinement fusion, combustion systems, and high-energy-density plasma dynamics \citep{sharp1984overview, zhou2017a_rayleigh}. The temporal evolution of RTI is generally characterized by three successive stages. Initially, small perturbations undergo exponential amplification governed by linear instability theory \citep{chandrasekhar1961hydrodynamic}. This is followed by a nonlinear regime characterized by the development of bubble-spike structures and coherent vortical roll-up. At sufficiently late times, the instability transitions into a turbulent mixing regime involving strong interfacial deformation, multiscale vortex interaction, and self-similar growth of the mixing layer \citep{youngs1984numerical}. A key parameter governing RTI evolution is the Atwood number,

\begin{equation}
At = \frac{\rho_1-\rho_2}{\rho_1+\rho_2},
\end{equation}

\noindent where $\rho_1$ and $\rho_2$ denote the densities of the heavier and lighter fluids, respectively. The Atwood number determines the strength of the buoyancy forcing and therefore strongly influences instability growth, vortex dynamics, turbulent mixing, and energy redistribution.

Considerable effort has been devoted toward understanding the role of Atwood number in single-interface RTI. Classical experimental investigations \citep{read1984experimental, andrews1990simple, roberts2016effects} and numerical simulations \citep{youngs1984numerical, youngs1989modelling, parhi1991sufficient, sengupta2023role} established the dependence of mixing-layer growth on Atwood number and demonstrated the emergence of self-similar turbulent evolution at late times. Subsequent experimental and numerical studies further quantified the influence of density gradient on bubble and spike dynamics, growth-rate coefficients, turbulence production, and spectral behaviour \citep{dimonte2004comparative, cook2004mixing, cabot2006reynolds, sengupta2022three}. More recent high-resolution simulations have shown that increasing Atwood number enhances baroclinic vorticity generation, accelerates nonlinear transition, and broadens the inertial range of the resulting turbulent cascade \citep{zhou2017a_rayleigh, zhou2017b_rayleigh}. Despite the substantial progress achieved for classical two-fluid RTI, the overwhelming majority of existing studies continue to focus on single-interface configurations.

In contrast, many naturally occurring and engineering flows involve multiple unstable density interfaces evolving simultaneously. Such multi-layer RTI systems arise in atmospheric and oceanic stratification \citep{tremblin2021radiative}, estuarine flows with overlapping salinity and thermal gradients, combustion and reacting-flow systems \citep{tavares2024dynamics}, astrophysical convection, and inertial confinement fusion capsules \citep{zhou2017a_rayleigh}. In these configurations, the instability evolution is governed not only by local density gradients but also by nonlinear coupling between adjacent interfaces through pressure perturbations, induced velocity fields, and turbulent entrainment \citep{mikaelian1982normal, sundaram2022non}. Consequently, the growth of one interface may either enhance or suppress the evolution of neighbouring interfaces depending on the stratification strength and layer separation.

Early theoretical investigations of multi-layer RTI by \citet{mikaelian1982rayleigh} demonstrated that perturbation growth in multi-interface systems is intrinsically coupled and depends on the complete density stratification rather than on local Atwood numbers alone. Subsequent analytical studies examined the interaction and merging behaviour of adjacent mixing layers in stratified systems \citep{parhi1991sufficient, yang1993general}. It was observed that depending on the separation distance (layer thickness) and density ratios, interfaces can either evolve quasi-independently or undergo a \lq \lq merging transition" to form a unified, high-volume mixing zone. This transition significantly alters the transport of mass and momentum, as the interaction between interfaces can either enhance or suppress the growth of specific spectral modes compared to classical two-fluid models. Experimental investigations of multi-layer RTI remain comparatively limited owing to the difficulty of maintaining controlled multi-interface stratification, although notable studies include miscible three-layer experiments \cite{jacobs2005rayleigh}, magnetic-levitation-based configurations \cite{adkins2017interface}, analytical studies in cylindrical domains \citep{guo2017rayleigh}, and more recent geometrically constrained multi-layer systems \citep{Suchandra_Ranjan_2023, liang2024experimental3d}. The experimentalists found that bubble evolution is governed by wall-induced spatial constraints while spike growth is dominated by initial tip curvature. Recent advances in direct numerical simulations and compressible multiphase modelling have enabled more detailed investigation of coupled multi-layer dynamics. Sengupta {\it et al.} \cite{sengupta2023multi} examined compressible enstrophy budget and mixing-layer evolution in thermally stratified multi-layer RTI, while \citet{Suchandra_Ranjan_2023} established global energy budgets in experimental three-layer gas systems. More recently, researchers \cite{ferguson2024dns, USTUN2025134644, sengupta2021role} investigated species transport, compressibility effects, and coupled instability dynamics in multi-layer buoyancy-driven flows. 

Despite these recent advances, several important questions regarding multi-layer RTI remain unresolved. In particular, the transition from weakly coupled interface-localized instability growth to fully coupled multi-layer turbulent mixing remains poorly characterized in the low-Atwood regime ($At < 0.1$). Existing studies have not systematically examined how independent variation of adjacent interfacial density contrasts modifies the coupling between mixing layers, the redistribution of turbulent kinetic energy, and the partitioning of released gravitational potential energy between turbulent production and viscous dissipation. Furthermore, the spectral characteristics of fully developed multi-layer RTI turbulence and their dependence on interfacial Atwood number remain largely unexplored.

The present work addresses these issues through fully compressible three-dimensional direct numerical simulations of thermally stratified multi-layer RTI involving three layers of air heated differentially and two unstable density interfaces. Such a configuration closely mimics several practical combustion and reacting-flow environments in which the same working fluid develops localized density stratification due to spatially non-uniform heat release. In turbulent combustion systems, localized flame fronts generate regions of hot, low-density combustion products adjacent to relatively colder, denser reactants, thereby producing multiple unstable density interfaces within the same fluid medium \cite{tavares2024dynamics}. Consequently, the present thermally stratified air configuration provides a physically representative framework for isolating and examining the fundamental instability and turbulence mechanisms associated with buoyancy-driven multi-layer mixing in combustion-driven environments without the additional complexity introduced by multi-species chemistry or variable composition effects. Furthermore, a systematic parametric investigation is performed by independently varying the interfacial Atwood numbers over the range $0.008 \le At\le 0.063$. The evolution of the coupled instability dynamics is characterized using vorticity amplification, mixing-layer growth, density fluctuations, turbulent kinetic energy, and energy partitioning between gravitational potential energy, turbulent kinetic energy, and dissipation.

\section{Problem Formulation of the Multi-Layer RTI}\label{sec2}
\begin{figure}
  \centerline{\includegraphics[width=0.65\textwidth]{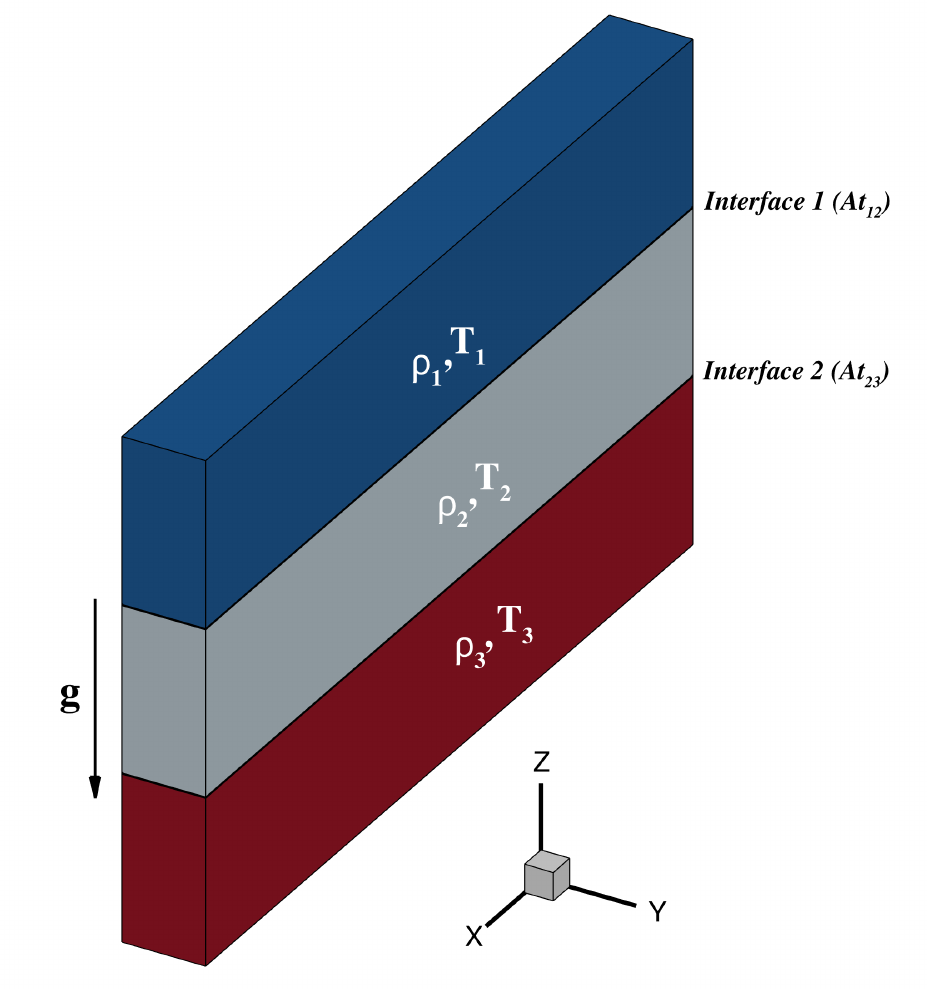}}
  \caption{ Schematic representation of the thermally stratified multi-layer Rayleigh-Taylor instability configuration, showing three layers separated by two interfaces.}
  \label{fig1}
\end{figure}

The multi-layer RTI simulations are performed in a three-dimensional rectangular domain, as shown in Fig. \ref{fig1}. Initially, the domain contains air arranged in multiple horizontal layers of uniform but distinct temperatures, and hence different densities, separated by insulating partitions at $t=0$. All bounding walls are adiabatically insulated, ensuring that the system remains thermodynamically isolated throughout the evolution. 

In the present numerical investigation, the multi-layer RTI and subsequent mixing are initiated at $t = 0$ by the impulsive horizontal removal of the separating partitions. Rather than imposing explicit artificial or external perturbations at the interfaces, this partition removal acts as the primary initial disturbance, triggering instability across a spectrum of spatial and temporal scales driven by a destabilizing potential energy. This approach mimics unforced experimental setups for hydrodynamic instabilities \citep{read1984experimental, roberts2016effects}. A more comprehensive characterization of the role of the initial perturbation on the subsequent RTI dynamics have been explored in a prior work of the research group \citep{sengupta2015effects}. Here, we use the same unforced partition removal as reported in experiments \cite{read1984experimental, roberts2016effects} with the objective of exploring evolution of a multi-layer stratified system subject to different interfacial Atwood numbers.

We resolve the unsteady 3D compressible Navier-Stokes equations (NSE) numerically, conserving mass, momentum, and energy.  Even though the Atwood numbers used across all the test cases reported here are small in magnitude, we have not used the Boussinesq approximation as Mikaelian \citep{Mikaelian2014} reported an underprediction of bubbles (fluid structures of light fluid growing into heavier fluid) by about 14\%, when RTI was formulated with the Boussinesq approximation. The flow is described by a set of five nonlinear partial differential equations, expressed in the divergence form \cite{sengupta2019direct, sengupta2019tracking, pal2026direct} as follows:

\begin{equation}
\frac{\partial \hat{Q}}{\partial t^*}+\frac{\partial \hat{E_c}}{\partial x^*}+\frac{\partial \hat{F_c}}{\partial y^*}+\frac{\partial \hat{G_c}}{\partial z^*}=\frac{\partial \hat{E_v}}{\partial x^*}+\frac{\partial \hat{F_v}}{\partial y^*}+\frac{\partial \hat{G_v}}{\partial z^*} + \hat{S},
\label{ge1}
\end{equation}

\noindent with the conserved variables given as

\begin{equation}
\hat{Q} = [ \rho^*; \; \rho^* u^*; \; \rho^* v^*; \; \rho^* w^*; \; \rho^* e_t^* ]^T.
\label{ge2}
\end{equation}

\noindent The convective flux variables $\hat{E}_c$, $\hat{F}_c$ and $\hat{G}_c$ are given as follows:

\begin{equation}
\hat{E}_c = [ \rho^* u^*; \; \rho^* u^{*2} + p^*; \; \rho^* u^* v^* ; \; \rho^* u^* w^*; \; (\rho^* e_t^* + p^*) u^*]^T,
\label{ge3}
\end{equation}

\begin{equation}
\hat{F}_c = [ \rho^* v^*; \; \rho^* u^* v^*; \; \rho^* v^{*2} + p^*; \; \rho^* v^* w^*; \; 
(\rho^* e_t^* + p^*) v^* ]^T,
\label{ge4}
\end{equation}

\begin{equation}
\hat{G}_c = [ \rho^* w^*; \; \rho^* u^* w^*; \; \rho^* v^* w^*; \; \rho^* w^{*2} + p^*; \;
(\rho^* e_t^* + p^*) w^* ]^T,
\label{ge5}
\end{equation}

\noindent and the viscous flux vectors $\hat{E}_v$, $\hat{F}_v$ and $\hat{G}_v$ are given as

\begin{equation}
\hat{E}_v = [ 0; \; \tau^*_{xx}; \; \tau^*_{xy}; \; \tau^*_{xz}; \; u^* \tau^*_{xx} + v^* \tau^*_{xy} + w^* \tau^*_{xz} - q^*_x ]^T,
\label{ge6}
\end{equation}

\begin{equation}
\hat{F}_v = [ 0; \; \tau^*_{yx}; \; \tau^*_{yy}; \; \tau^*_{yz}; \; u^* \tau^*_{yx} + v^* \tau^*_{yy} + w^* \tau^*_{yz} - q^*_y ]^T,
\label{ge7}
\end{equation}

\begin{equation}
\hat{G}_v = [ 0; \; \tau^*_{zx}; \; \tau^*_{zy}; \; \tau^*_{zz}; \; u^* \tau^*_{zx} + v^* \tau^*_{zy} + w^* \tau^*_{zz} - q^*_z ]^T.
\label{ge8}
\end{equation}

\noindent The source term, $\hat{S}$ is given by
\begin{equation}
\hat{S} = [ 0; \; 0; \; -\rho^*g; \; 0; \; -\rho^* v^* g ]^T
\end{equation}

\noindent The flow variables $\rho^*$, $p^*$, $u^*$, $v^*$, $w^*$, $T^*$ and $e_t^*$ represent dimensional density, pressure, Cartesian components of velocity, the absolute temperature and specific internal energy, respectively. The specific heat ratio, $\gamma$, is set to 1.4. The components of symmetric viscous stress tensor are $\tau^*_{xx}$, $\tau^*_{yy}$, $\tau^*_{zz}$, $\tau^*_{xy}$, $\tau^*_{xz}$, $\tau^*_{yx}$, $\tau^*_{yz}$, $\tau^*_{zx}$ and $\tau^*_{zy}$ which are related to the gradients of velocity as follows:

\begin{equation}
\tau^*_{xx} = \biggl( 2 \mu^* \frac{\partial u^*}{\partial x^*} + \lambda^* \nabla^* \cdot \vec{V}^* \biggr ); \; \tau^*_{yy} = \biggl( 2 \mu^* \frac{\partial v^*}{\partial y^*} + \lambda^* \nabla^* \cdot \vec{V}^* \biggr ); \; \tau^*_{zz} = \biggl( 2 \mu^* \frac{\partial w^*}{\partial z^*} + \lambda^* \nabla^* \cdot \vec{V}^* \biggr ),
\label{ge9}
\end{equation}

\begin{equation}
\tau^*_{xy} = \tau^*_{yx} = \mu^* \biggl( \frac{\partial u^*}{\partial y^*} +  \frac{\partial v^*}{\partial x^*} \biggr ); \; \tau^*_{xz} = \tau^*_{zx} = \mu^* \biggl( \frac{\partial u^*}{\partial z^*} +  \frac{\partial w^*}{\partial x^*} \biggr ); \; \tau^*_{yz} = \tau^*_{zy} = \mu^* \biggl( \frac{\partial v^*}{\partial z^*} +  \frac{\partial w^*}{\partial y^*} \biggr ).
\label{ge10}
\end{equation}

\noindent In this formulation, the specific heat capacity ($C_v$) and thermal conductivity ($\kappa$) are considered as constants. The heat conduction terms $q^*_x$, $q^*_y$, and $q^*_z$ are expressed as follows:

\begin{equation}
q^*_x = - \kappa \frac{\partial T^*}{\partial x^*}, \; q^*_y = - \kappa \frac{\partial T^*}{\partial y^*}, \; q^*_z = - \kappa \frac{\partial T^*}{\partial z^*}.
\label{ge11}
\end{equation}

\noindent Stokes' hypothesis establishes a connection between molecular viscosity, $\mu^*$, and the second coefficient of viscosity, $\lambda^*$, as $\lambda^* = -2\mu^*/3$. Previous simulations of RTI \citep{sengupta2016roles, sengupta2016non} have demonstrated the necessity of incorporating non-zero values of bulk viscosity to accurately capture the early-time behavior. This is achieved by utilizing acoustic attenuation measurements of Ash {\it et al.} \cite{ash1991second} to determine the bulk viscosity of air. Sutherland's law is employed to calculate viscosity as a function of temperature. The ideal gas equation serves as the closure for the system of equations, yielding:

\begin{equation}
p^*=\rho^* R^*T^*,
\label{ge12}
\end{equation}

\noindent which in turn defines specific energy, $e_t^*$ as

\begin{equation}
e_t^* =\frac{p^*}{[\rho^*(\gamma-1)]} + \frac{(u^{*2} + v^{*2} + w^{*2})}{2}.
\label{ge13}
\end{equation}

\noindent The independent and dependent variables are nondimensionalized as follows

\begin{equation}
(x, y, z) = \biggl ( \frac{x^*}{L}, \frac{y^*}{L}, \frac{z^*}{L} \biggr ), \; (u, v, w) = \biggl ( \frac{u^*}{U_s}, \frac{v^*}{U_s},\frac{w^*}{U_s} \biggr ), \;t = \frac{t^*U_s}{L},
\label{ge14}
\end{equation}

\begin{equation}
\rho = \frac{\rho^*}{\rho_s}, \; \theta = \frac{T^* - T_s}{\Delta T}, \; p = \frac{p^* - p_s} {\rho_s U^2_s}, \; e_t = \frac{e^*_t - e_{ts}}{U^2_s}, \; \mu = \mu^*/\mu_s.
\label{ge15}
\end{equation}

\noindent The characteristic parameters in the study are expressed in terms of five dimensionless quantities \cite{kumar2026supervised, sengupta2025bifurcation}: Reynolds number ($Re$), Prandtl number ($Pr$), Froude number ($Fr$), Eckert number ($Ec$), and Gay-Lussac number ($Ga$). These parameters are defined with respect to reference values: length scale $L$ and velocity scale, $U_s$, and density scale $\rho_s$. To investigate the effect of thermal stratification on the evolution of a multi-fluid RTI, a series of numerical simulations is performed for a three-fluid configuration forming two distinct interfaces, as shown in Fig. \ref{fig1}. In all cases, the coldest fluid is placed at the top of the domain with a fixed temperature $T_{1} = 298~\mathrm{K}$, while the middle and lower fluids are maintained at progressively higher temperatures, $T_2$ and $T_3$. This results in an unstably stratified configuration satisfying $\rho_1 > \rho_2 > \rho_3$, leading to the development of RTI at both fluid interfaces. The temperature scale for all simulations is computed as $\Delta T = T_3 - T_1$. The reference condition corresponds to air at normal temperature and pressure. The chosen values for the velocity scale, $U_s$, and length scale, $L$, are 0.64539 m/s and 0.00963 m, respectively, resulting in $Re$ of 6944. The similitude of the flow is then described as follows:

\begin{equation}
Re = \frac{\rho_s U_s L}{\mu_s}, \; Pr = \frac{\mu C_p}{\kappa}, \; Fr = \sqrt{\frac{U^2_s}{gL}}, \; Ec = \frac{U^2_s}{C_p \Delta T}, \; Ga = \frac{\Delta T}{T_s},
\label{ge16}
\end{equation}

\noindent where the Gay-Lussac number relates temperature difference from a reference temperature ($T_s = T_1 = 298$ K). The reference density is taken as $\rho_s = 1.225$~$\rm kg/m^3$. In the present computations, $Pr = 0.712$ and $Fr = 0.144$. A total of ten test cases (A-J) are considered, wherein the temperature differences across the top and bottom interfaces are systematically varied. These temperature gradients give rise to two Gay-Lussac numbers, $\mathrm{Ga}_{12}$ and $\mathrm{Ga}_{23}$, corresponding to the top and bottom interfaces, respectively. The values of $\mathrm{Ga}_{12}$ and $\mathrm{Ga}_{23}$ increase monotonically from Case-A to Case-J, enabling a detailed parametric assessment of buoyancy-driven instability strength in a two-layer RTI system involving three fluids. In the present study, two Atwood numbers are defined: $\mathrm{At}_{12}$ for the top interface and $\mathrm{At}_{23}$ for the bottom interface. The corresponding values for all test cases are reported in Table~\ref{tab1}.

Thus, the nondimensionalized governing equations are 

\begin{equation}
\frac{\partial Q}{\partial t}+\frac{\partial E_c}{\partial x}+\frac{\partial F_c}{\partial y}+\frac{\partial G_c}{\partial z}=\frac{\partial E_v}{\partial x}+\frac{\partial F_v}{\partial y}+\frac{\partial G_v}{\partial z}.
\label{ge17}
\end{equation}

\noindent The viscous shear stresses with non-zero bulk viscosity use experimental measurements \citep{ash1991second}. This gives a measure of $\mu_b$ for air as a function of temperature ($T^*$ in K) using linear regression of the data as $\mu_b = -7.383 \times 10^{-4} + 3.381 \times 10^{-4} T^*$. The non-dimensional viscous stress components are evaluated as

\begin{equation*}
\tau_{xx} = \frac{1}{Re} \biggl [ \biggl ( \frac{4 \mu}{3} + \frac{\mu_b}{\mu_s} \biggr ) \frac{\partial u}{\partial x} + \biggl (-\frac{2 \mu}{3} + \frac{\mu_b}{\mu_s} \biggr ) \frac{\partial v}{\partial y} + \biggl (-\frac{2 \mu}{3} \mu + \frac{\mu_b}{\mu_s} \biggr ) \frac{\partial w}{\partial z} \biggr ], 
\end{equation*} 
\begin{equation*}
\tau_{yy} = \frac{1}{Re} \biggl [ \biggl ( \frac{4}{3} \mu + \frac{\mu_b}{\mu_s} \biggr ) \frac{\partial v}{\partial y} + \biggl (-\frac{2}{3} \mu + \frac{\mu_b}{\mu_s} \biggr ) \frac{\partial u}{\partial x} + \biggl (-\frac{2}{3} \mu + \frac{\mu_b}{\mu_s} \biggr ) \frac{\partial w}{\partial z} \biggr ],
\end{equation*}
\begin{equation}
\tau_{zz} = \frac{1}{Re} \biggl [ \biggl ( \frac{4}{3} \mu + \frac{\mu_b}{\mu_s} \biggr ) \frac{\partial w}{\partial z} + \biggl (-\frac{2}{3} \mu + \frac{\mu_b}{\mu_s} \biggr ) \frac{\partial u}{\partial x} + \biggl (-\frac{2}{3} \mu + \frac{\mu_b}{\mu_s} \biggr ) \frac{\partial v}{\partial y} \biggr ],
\label{ge18}
\end{equation}

\begin{equation}
\tau_{xy} = \tau_{yx} = \frac{\mu}{Re} \biggl [\frac{\partial u}{\partial y} + \frac{\partial v}{\partial x} \biggr ], \; \tau_{xz} = \tau_{zx} = \frac{\mu}{Re} \biggl [\frac{\partial u}{\partial z} + \frac{\partial w}{\partial x} \biggr ], \;\tau_{yz} = \tau_{zy} = \frac{\mu}{Re} \biggl [\frac{\partial w}{\partial y} + \frac{\partial v}{\partial z} \biggr ].
\label{ge19}
\end{equation}

\noindent The heat conduction terms are given as follows

\begin{equation}
q_x = - \frac{\mu}{Ga Re Pr Ec} \frac{\partial \theta}{\partial x}, \; q_y = - \frac{\mu}{Ga Re Pr Ec} \frac{\partial \theta}{\partial y}, \; q_z = - \frac{\mu}{Ga Re Pr Ec} \frac{\partial \theta}{\partial z}.
\label{ge20}
\end{equation}

\noindent The equation of state in nondimensional form is given by $p = \rho R \theta - \frac{R}{Ga} (1- \rho)$, where $R = R^* \Delta T / U_s^2$. Here, we use a uniform Cartesian grid for the computational domain, as depicted in Fig. ~\ref{fig1}. Parallelization is achieved using Schwartz domain decomposition with eight overlap points across sub-domain boundaries.

\subsection{Numerical methodology}

\noindent The current numerical investigation utilizes a computational domain displayed in Fig. \ref{fig1}. The calculations are conducted on a 3D mesh, with the domain being a cuboid with dimensions $0 \le x \le 0.245$, $0 \le y \le 0.15$, and $0 \le z \le 0.025$.  Figure~\ref{fig1} presents a schematic representation of the thermally stratified, two-interface Rayleigh-Taylor instability (RTI) configuration investigated in the present study. The system consists of a single working fluid (air) arranged in three miscible layers with densities $\rho_1$, $\rho_2$, and $\rho_3$ from top to bottom such that $\rho_1 > \rho_2 > \rho_3$, forming a gravitationally unstable stratification under the action of downward gravity $g$. Density stratification is established via initial step profiles in temperature governed by the ideal gas equation of state. The two unstable interfaces separating the layers define distinct RTI-prone regions characterized by independent Atwood numbers, $At_{12}$ and $At_{23}$, corresponding to the upper (interface-1) and lower (interface-2) interfaces, respectively. The Cartesian coordinate system employed in the numerical simulations is indicated, with the vertical direction $y$ aligned with gravity. This schematic defines the baseline physical configuration used for the parametric investigation of multi-layer RTI presented in this work.

The boundary conditions for the primitive variables $u$, $v$, $w$, $p$, and $\theta$ are obtained from the conserved variables $\hat{Q}$ at each time-step. On the walls of the 3D box where $x \ge 0$ and $y \ge 0$, no-slip adiabatic conditions are imposed. The thermodynamically isolated system, utilizing the current compressible formulation, provides a solution field to study RTI in all its essential details, from the initial linear stage to subsequent nonlinear ones. The corresponding non-dimensional wall boundary conditions are as follows:

\begin{equation}
u = 0, \quad v = 0, \quad w = 0, \quad \frac{\partial \theta}{\partial n} = 0, \quad \frac{\partial p}{\partial n} = 0.
\label{ge23}
\end{equation}

At the beginning of each simulation ($t = 0$), the heavier air on top is assigned a temperature of $T = T_s$, resulting in $\theta = 0$ for the coldest (top) air and $\theta = 1$ for the warmest air at the bottom. The middle layer will have values of $\theta$ ranging from 0.25 (case-A) to 0.4375 (case-J) depending on the temperature differential recorded in Table \ref{tab1}. By using air at different temperatures to simulate the density differential occurring in RTI, there is no need for interface tracking of two different fluids. Note that the governing equations are coupled with the ideal gas law \cite{sengupta2024separation, sengupta2024thermal}, given in Eq. \eqref{ge12}. This follows the approach adopted in prior studies \citep{sengupta2021role, ajanif2026multi}, which examined all essential features of RTI from onset to the turbulent mixing stage. The thermal boundary condition indicates that there is no heat transferred to or from the 3D box to the surroundings, while the pressure boundary condition considers the boundary layer assumption at the wall. In the multi-layer systems, the heaviest (coldest) air is located at the top of the domain, with the two interfaces initially being placed at $y = 0.05$ and $0.1$, respectively.

\begin{table}[!ht]
\caption{Summary of numerical parameters and corresponding test case descriptions}
\label{tab1}
\label{CaseHistory}
\begin{ruledtabular}
\begin{tabular}{c c c c c c c}
\textbf{Case} & \textbf{$T_{\mathrm{upper}}$ (K)} & \textbf{$T_{\mathrm{middle}}$ (K)} & \textbf{$T_{\mathrm{lower}}$ (K)} & \textbf{At$_{12}$} & \textbf{At$_{23}$} & \begin{tabular}[c]{@{}c@{}}\textbf{Final simulation}\\\textbf{time}\end{tabular} \\
\hline
A & 298 & 303   & 318 & 0.0083 & 0.0245 & 120.0 \\
B & 298 & 305.5 & 323 & 0.0124 & 0.0278 & 89.0  \\
C & 298 & 308   & 328 & 0.0165 & 0.0324 & 64.8  \\
D & 298 & 310.5 & 333 & 0.0205 & 0.0350 & 56.5  \\
E & 298 & 313   & 338 & 0.0245 & 0.0403 & 50.0  \\
F & 298 & 315.5 & 343 & 0.0285 & 0.0418 & 40.0  \\
G & 298 & 318   & 348 & 0.0325 & 0.0450 & 37.8  \\
H & 298 & 323   & 358 & 0.0402 & 0.0458 & 37.3  \\
I & 298 & 328   & 368 & 0.0479 & 0.0575 & 31.5  \\
J & 298 & 333   & 378 & 0.0555 & 0.0633 & 29.9 \\
\end{tabular}
\end{ruledtabular}
\end{table}

Ten numerical simulations are reported in the present study, the details of which are summarized in Table~\ref{tab1}. The simulations correspond to a two-interface, three-layer stratified configuration, in which the density gradients across successive interfaces is varied systematically. The computations involved require 2.36 million core hours of computing. The cases span a wide range of stratification strengths, characterized by the corresponding Atwood numbers, $At_{12}$ and $At_{23}$ across the upper and lower interfaces, respectively. Specifically, the Atwood number across the upper interface varies from $At_{12}=0.008$ to $0.056$, while that across the lower interface varies from $At_{23}=0.025$ to $0.063$, thereby allowing a systematic assessment of the influence of increasing density stratification on the growth and interaction of multiple mixing layers.

To maintain spatial discretization consistency across the parametric study, all ten simulation cases (Cases A-J) are computed on an identical, fine 3D structured mesh consisting of $1066 \times 640 \times 106$ grid points in the $x$-, $y$-, and $z$-directions, respectively. All simulations are performed using the same numerical framework and grid resolution adopted in the earlier investigations of single-layer RTI\cite{sengupta2021role}. Extensive experimental validation studies were carried out in these references, where an accurate quantitative agreement of the mixing layer heights and growth rate coefficients was obtained with the experiments of Read \cite{read1984experimental}. The qualitative contour plots of density also showed a milli-second level match with the experimental frames \cite{read1984experimental}. The same validated computational setup is employed here for the present multi-layer RTI simulations.

As all cases utilize the exact same domain geometry, numerical scheme, time step ($1 \times 10^{-5}$), and spatial grid resolution ($1066 \times 640 \times 106$), the numerical discretization framework is held strictly uniform across all simulations. Under these controlled conditions, the initial baroclinic vorticity generated at the sidewalls ($\nabla \rho \times \nabla p$) is governed directly by the physical density difference ($\Delta \rho$) of each specific case, rather than spatial grid resolution artifacts. Consequently, all observed relative variations in onset times, growth rates, and flow topologies across Cases A-J are driven purely by physical stratification effects.

The computations in this study utilize dispersion-relation-preserving, highly accurate compact schemes for convective derivative calculations \citep{sagaut2023global, sengupta2023compressibility, joshi2023exploring}. This ensures that the numerical and physical dispersion relations match over a broad range of spatial and temporal scales, which is crucial for accurately capturing RTI. This is especially important during the onset from the sidewalls of the 3D box \cite{andrews1990simple}. A compact three-dimensional filter with a filter coefficient of 0.15 and a frequency of 50 is employed. For evaluating the convective flux derivatives, a high-resolution optimized upwind compact scheme (OUCS3) \citep{sagaut2023global, sengupta2025benchmark, sengupta2020effects} is used. The second-order central difference ($CD_2$) scheme is utilized for viscous flux calculations, and time integration is performed using the OCRK3 scheme \citep{sagaut2023global, sengupta2025compressible, sengupta2020effectsa}, which is an optimized version of the three-stage Runge-Kutta method. The time-step prescribed for the simulations is $1 \times 10^{-5}$, resulting in CFL numbers in the $x$-, $y$-, and $z$-directions of 0.0692, 0.04265, and 0.0344, respectively.

\section{RESULTS AND DISCUSSIONS}

This section presents the instability evolution and turbulent mixing dynamics in thermally stratified multi-layer Rayleigh-Taylor instability for a broad range of interfacial density gradients. The temporal evolution of the flow is first examined using the maximum vorticity and characteristic instability timescales associated with the onset, exponential growth, and nonlinear saturation regimes. The corresponding vortical structures and mixing-layer development are subsequently analysed through vorticity contours at representative stages of the flow evolution. Quantitative characterization of the instability growth is then performed using the temporal evolution of the individual and merged mixing-layer heights together with the associated growth-rate coefficients.

The later part of the section focuses on the wall-normal distributions of density fluctuations and turbulent kinetic energy at different stages of instability development in order to characterize the evolution of mixing and turbulence redistribution across the stratified layers. Finally, spectral analysis of the late-stage vorticity at the central wall-normal location is performed using Fourier transforms to examine the wavenumber scaling of the fully developed flow.

\subsection{Vorticity dynamics of the multi-layer RTI}

In this section, the spatio-temporal evolution of the multi-layer RTI is examined using instantaneous vorticity contours corresponding to three critical regimes identified in the temporal distribution of maximum vorticity. These stages together provide a coherent description of the transition from interfacial instability to fully developed turbulent mixing. 

\subsubsection{Onset, exponential growth and nonlinear saturation stages of RTI}

\begin{figure}
  \centerline{\includegraphics[width=0.90\textwidth]{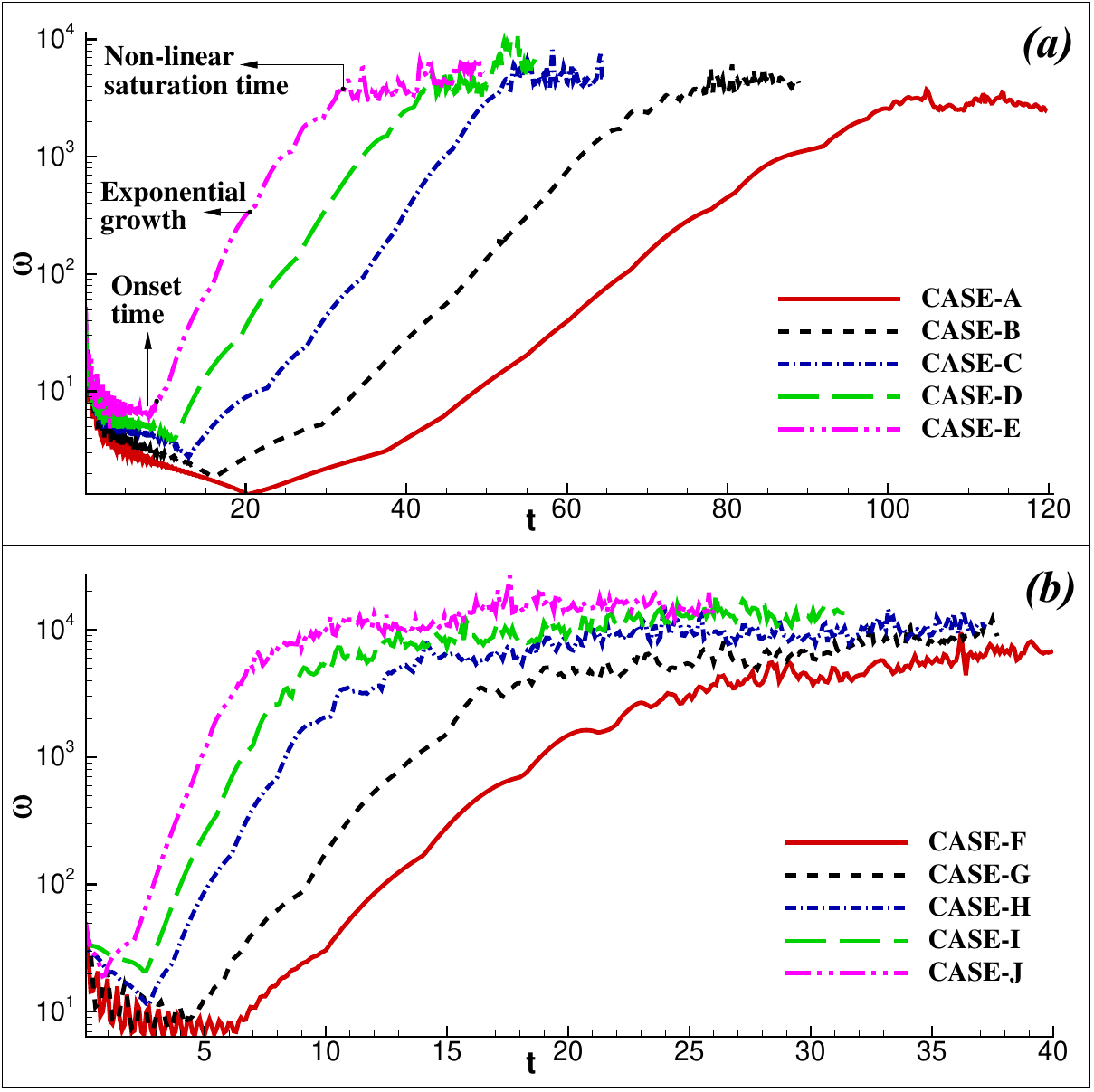}}
  \caption{Time evolution of maximum vorticity for all simulated cases: ($a$) cases A-E and ($b$) cases F-J.}
  \label{fig2}
\end{figure}

Figure \ref{fig2} presents the temporal evolution of the maximum vorticity magnitude ($\omega_{max}$), for all simulated configurations. Cases A-E and F-J are shown separately in the upper and lower panels, respectively, in increasing order of interfacial $At$. The evolution of $\omega_{max}$ exhibits three distinct stages common to all cases: an initial onset regime, a quasi-exponential growth regime, and a nonlinear saturation regime.

During the initial stage, $\omega_{max}$ undergoes a slight decay before increasing monotonically. This transient behaviour arises from the initial dominance of viscous diffusion over baroclinic vorticity generation during the early adjustment of the stratified interfaces \citep{sengupta2022three}. Although the density gradients are imposed at $t=0$, the interfacial deformation remains weak during the initial stages, resulting in limited misalignment between the pressure and density gradients and consequently weak baroclinic torque production. The onset time is therefore identified as the instant beyond which $\omega_{max}$ begins to increase monotonically owing to sustained buoyancy-driven vorticity amplification.

Following the onset stage, all cases enter a quasi-exponential growth regime characterized by an approximately linear variation of $log(\omega_{max})$ with time. This phase corresponds to rapid amplification of interfacial vorticity driven predominantly by baroclinic torque associated with the unstable density stratification \citep{sharp1984overview, dimonte2004comparative}. Increasing interfacial $At$ leads to progressively earlier onset and steeper growth of $\omega_{max}$, indicating enhanced buoyancy forcing and stronger vorticity production at the interfaces. In contrast, the weakly stratified configurations exhibit delayed amplification and comparatively slower growth rates, consistent with reduced baroclinic activity.

At later times, the growth of $\omega_{max}$ departs from the quasi-exponential regime and transitions into a nonlinear saturation stage marked by strong temporal fluctuations and reduced mean growth. This regime corresponds to the onset of intense vortex interaction, coherent-structure breakdown, and redistribution of vorticity across a broad range of spatial scales as the flow evolves toward fully developed turbulent mixing \citep{goncharov2002analytical, boffetta2017incompressible}. The cases with higher $At$ transition more rapidly into the nonlinear regime and attain substantially larger peak vorticity levels compared to the weakly stratified configurations, indicating accelerated instability development and enhanced turbulent mixing dynamics.

\begin{figure*}[htbp]
  \centering
  \includegraphics[width=0.95\textwidth]{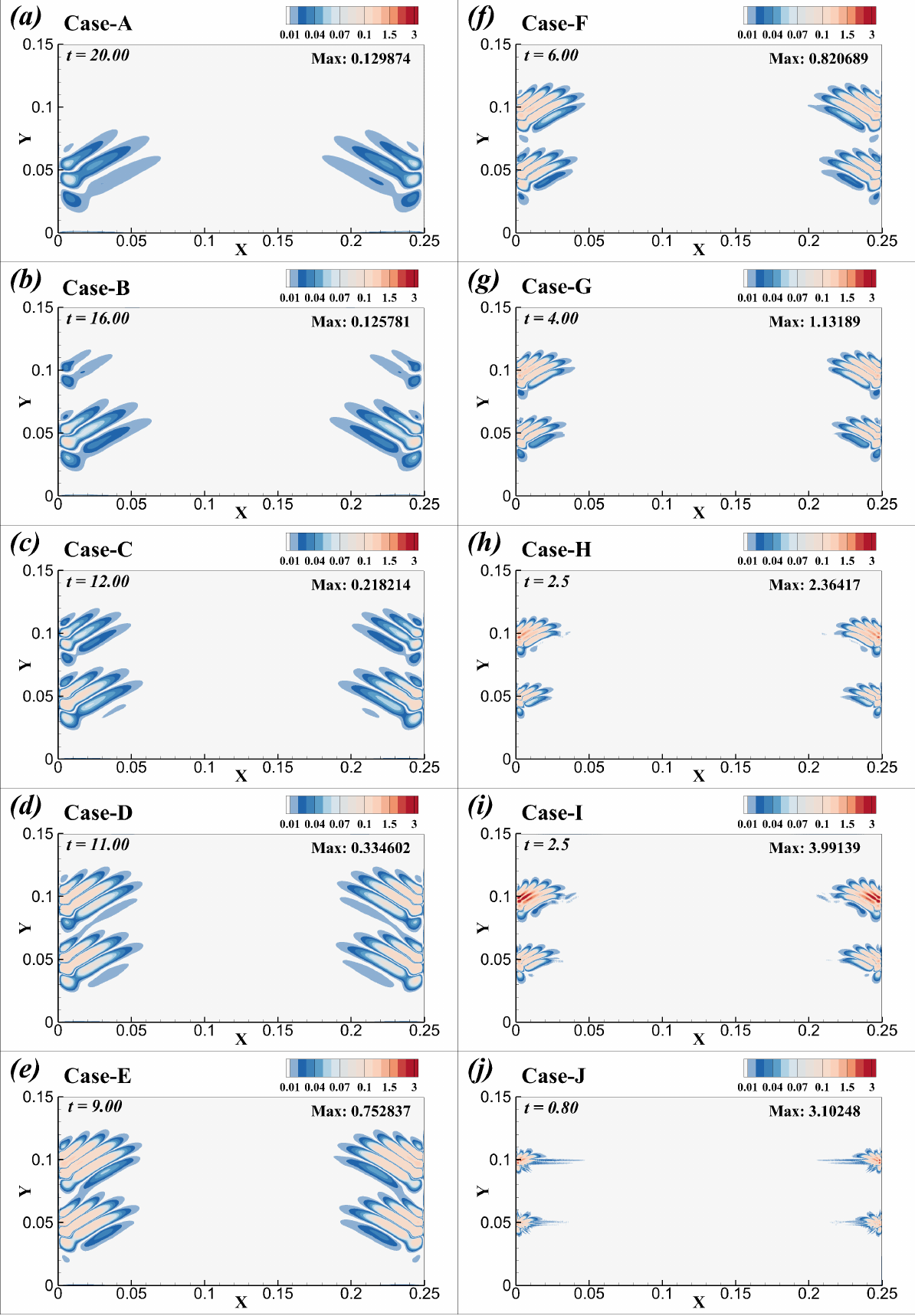}
  \caption{Vorticity contours during the onset stage for all cases.}
  \label{fig3}
\end{figure*}

Figure \ref{fig3} presents the vorticity contours at the instability onset times identified in Fig. \ref{fig2} across all investigated configurations. At this early stage, vorticity generation is localized within thin regions surrounding the two unstable density interfaces at $y \approx 0.05$ and $y \approx 0.10$, confirming that the flow is in the initial baroclinic growth regime. The observed vortical structures arise from baroclinic torque generated by the spatial misalignment of pressure and density gradients ($\nabla \rho \times \nabla p$) initiated near the lateral boundaries, which serves as the primary mechanism governing early-stage RTI development \citep{sharp1984overview, cook2004mixing}.

The onset-stage vortical structures remain spatially coherent and symmetric, initiating near the lateral boundaries ($x = 0$ and $x = 0.25$) and propagating inward toward the center of the domain. The spatial distribution and intensity of the vorticity deposition display a strong dependency on the stratification regime. In the lower-Atwood regime (Cases A--E), the lower interface ($y \approx 0.05$) experiences broader spatial dispersion over a wider vertical extent due to the higher local density ratio ($At_{23} > At_{12}$). Conversely, in the higher-Atwood regime (Cases F--J), enhanced thermal gradients concentrate the baroclinic production into tightly confined vortex cores along the upper interface ($y \approx 0.10$), leading to high localized peak vorticity values (represented by the dark red contours in Cases I and J). Across all configurations, increasing the overall density stratification drastically reduces the onset time (from $t = 20.0$ in Case A to $t = 0.80$ in Case J), demonstrating that stronger buoyancy forcing accelerates early baroclinic vorticity deposition and accelerates the onset of multi-layer RTI dynamics.

\begin{figure*}[htbp]
  \centering
  \includegraphics[width=0.92\textwidth]{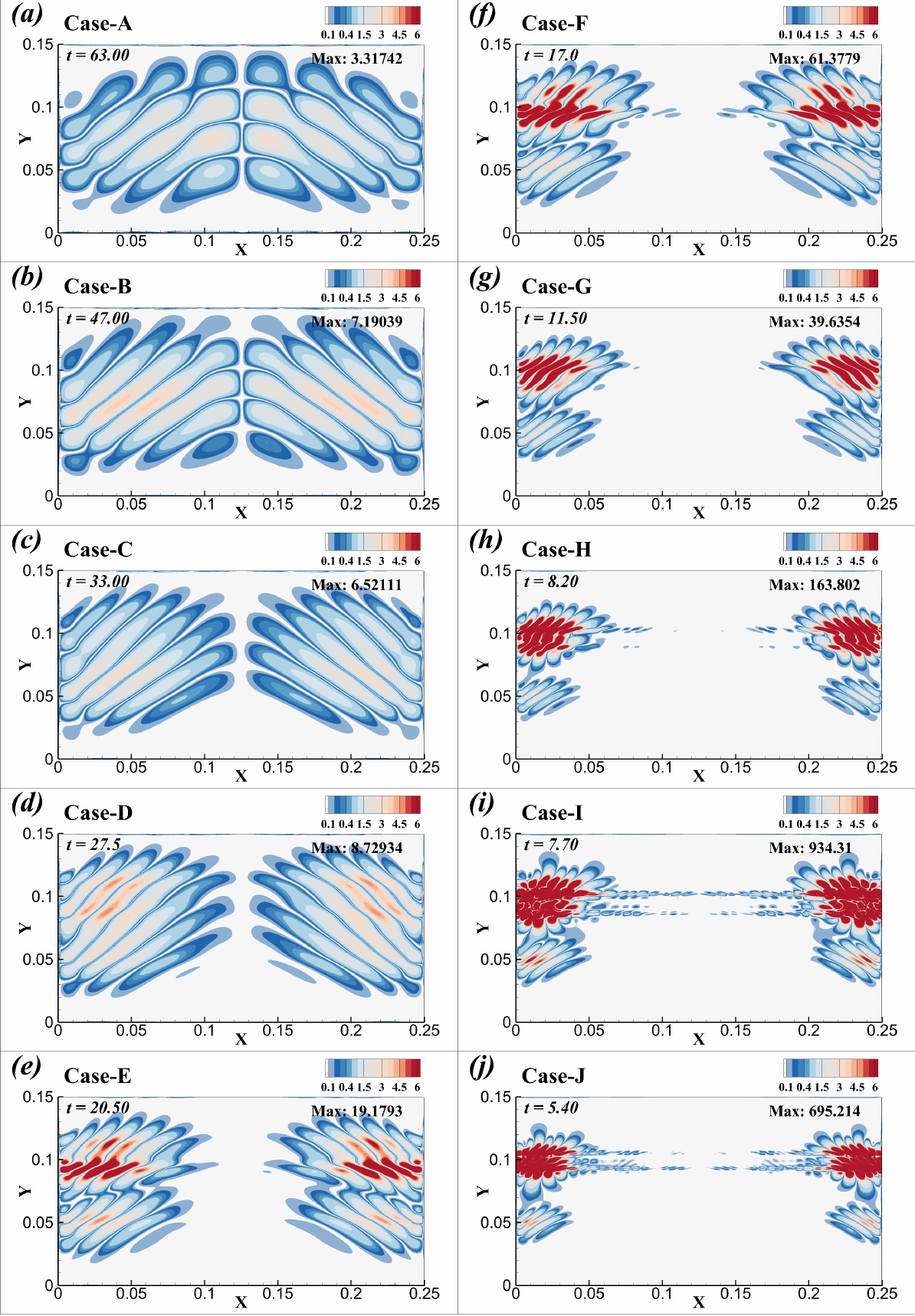}
  \caption{Vorticity contours during the log-linear growth stage for all cases.}
  \label{fig4}
\end{figure*}

Figure \ref{fig4} presents the vorticity contours at a characteristic intermediate time, $t^*$, defined as the arithmetic mean of the onset time and the time corresponding to the start of nonlinear saturation. This time instance lies exactly within the quasi-exponential (log-linear) growth regime identified in Fig. \ref{fig2}, enabling a direct comparison of how the early onset structures observed in Fig. \ref{fig3} evolve prior to fully nonlinear saturation.

As time progresses from the onset stage (Fig.\ref{fig3}) to the log-linear regime (Fig. \ref{fig4}), the flow topology exhibits a pronounced transition dictated by the stratification strength. In the weakly stratified regime (Cases A--D), the localized boundary disturbances initially seen in Fig. \ref{fig3} propagate inward toward the domain center ($x = 0.125$) and expand vertically. This leads to domain-spanning, strongly coupled structures where vortical structures extend continuously across both interfaces ($y \approx 0.05$ and $y \approx 0.10$), filling nearly the entire vertical domain ($y \in [0.02, 0.14]$). 
Case E marks a transitional phase where intense vorticity begins to concentrate along the upper interface ($y \approx 0.10$), with maximum vorticity jumping to $\omega_{\max} \approx 19.18$. For the strongly stratified configurations (Cases F--J), the early onset structures do not merge across the domain height; instead, they develop into two spatially decoupled horizontal mixing bands. The upper band ($y \approx 0.10$) experiences significantly stronger local baroclinic torque ($\nabla \rho \times \nabla p$), rapidly amplifying the initial boundary perturbations into compact, high-magnitude vorticity cores (reaching $\omega_{\max} > 900$ in Case I). Meanwhile, vorticity along the lower band ($y \approx 0.05$) remains comparatively diffuse and spatially spread, confirming that higher density stratification accelerates the growth rate and sharpens the spatial decoupling between the two mixing layers \citep{cook2004mixing}.

\begin{figure*}
  \centerline{\includegraphics[width=0.95\textwidth]{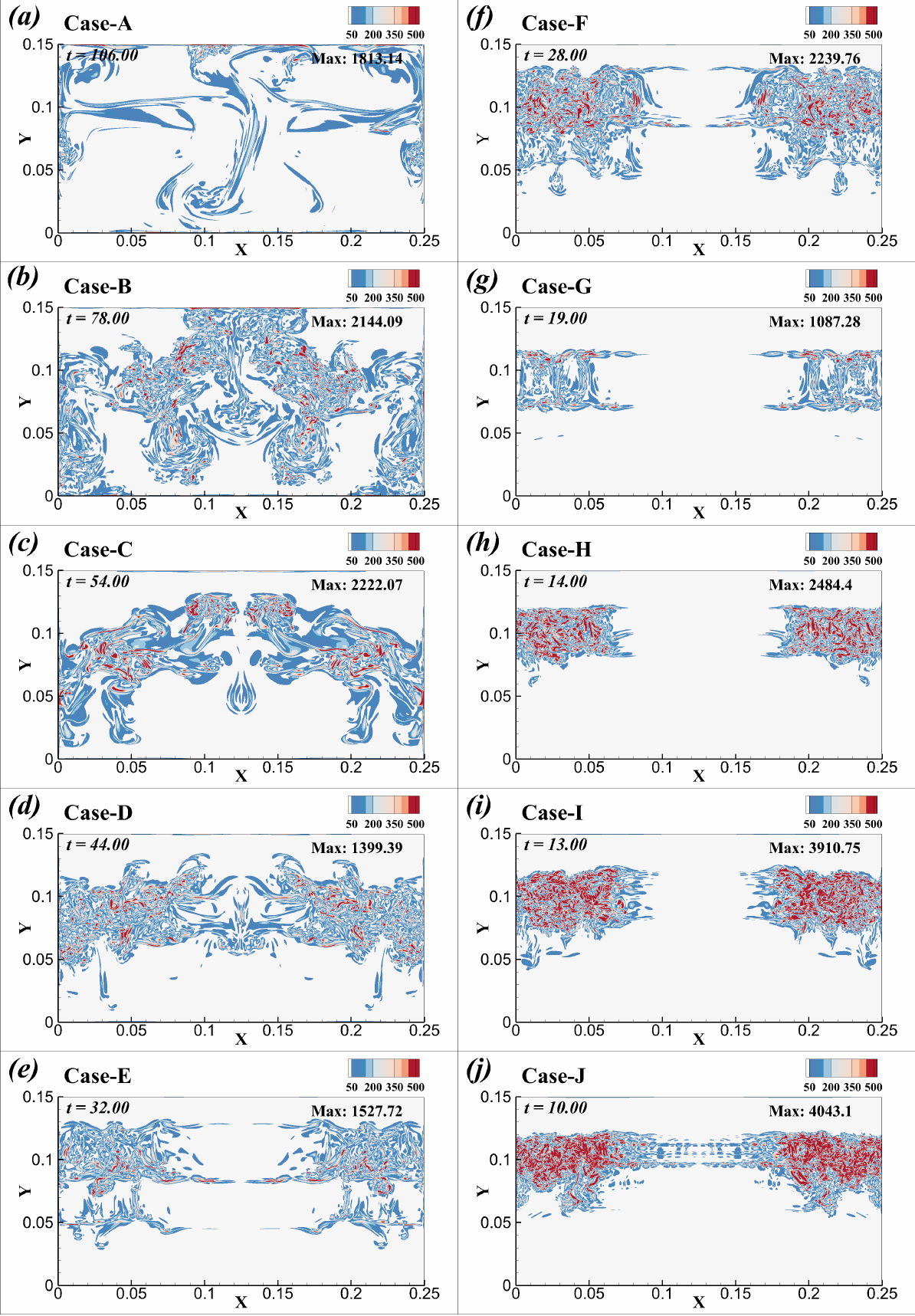}}
  \caption{Vorticity contours during the non-linear saturation stage for all cases.}
  \label{fig5}
\end{figure*}

Figure~\ref{fig5} presents the vorticity contours during the nonlinear saturation stage for all investigated configurations. The selected snapshots correspond to the onset of saturation identified in Fig. \ref{fig2}, at which the rapid quasi-exponential amplification of $\omega_{max}$ transitions into a regime dominated by nonlinear vortex interaction and turbulent redistribution of vorticity.

The weakly stratified configurations (Cases A and B) exhibit broad, spatially distributed vortical structures occupying a substantial fraction of the computational domain. The initially coherent bubble-spike structures observed during the exponential growth stage undergo progressive distortion and fragmentation, producing irregular large-scale eddies characteristic of nonlinear single-interface RTI evolution \citep{sharp1984overview, dimonte2004comparative}. In these cases, the mixing dynamics remain dominated primarily by the lower interface, consistent with the effectively single-interface behaviour identified previously. For the intermediate stratification cases (Cases C-E), the vortical field becomes increasingly multiscale owing to enhanced interaction between structures generated at the two interfaces. The initially separated mixing layers broaden substantially and develop pronounced interfacial coupling, leading to fragmentation of the coherent vortices into smaller interacting eddies. Case E in particular exhibits clear evidence of simultaneous upper- and lower-interface activity, marking the transition toward fully coupled multi-layer dynamics. The strongly stratified configurations (Cases F-J) display a qualitatively different flow topology characterized by highly localized regions of intense vorticity concentrated near the unstable interfaces. The stronger density gradients generate compact clusters of small-scale vortices associated with enhanced baroclinic production and interfacial shear-layer instability \citep{andrews1990simple}. In Cases I and J, thin elongated vortical structures extending toward the center of the domain indicate strong interaction between the two actively evolving mixing layers and the onset of layer merger. Despite the larger lower-interface density gradient, the upper mixing region exhibits stronger late-stage vortical activity, suggesting substantial nonlinear redistribution of vorticity during the saturation process.

Overall, the nonlinear saturation stage is characterized by the breakdown of the coherent structures established during the quasi-exponential growth regime and the emergence of strongly interacting multiscale vortical dynamics. Increasing interfacial stratification fundamentally alters the organization of the saturated flow by driving the transition from weakly coupled, effectively single-interface behaviour to fully coupled multi-layer turbulent mixing.

\subsubsection{Comparing vorticity evolution in weakly and strongly stratified regimes}

\begin{figure}
  \centerline{\includegraphics[width=\textwidth]{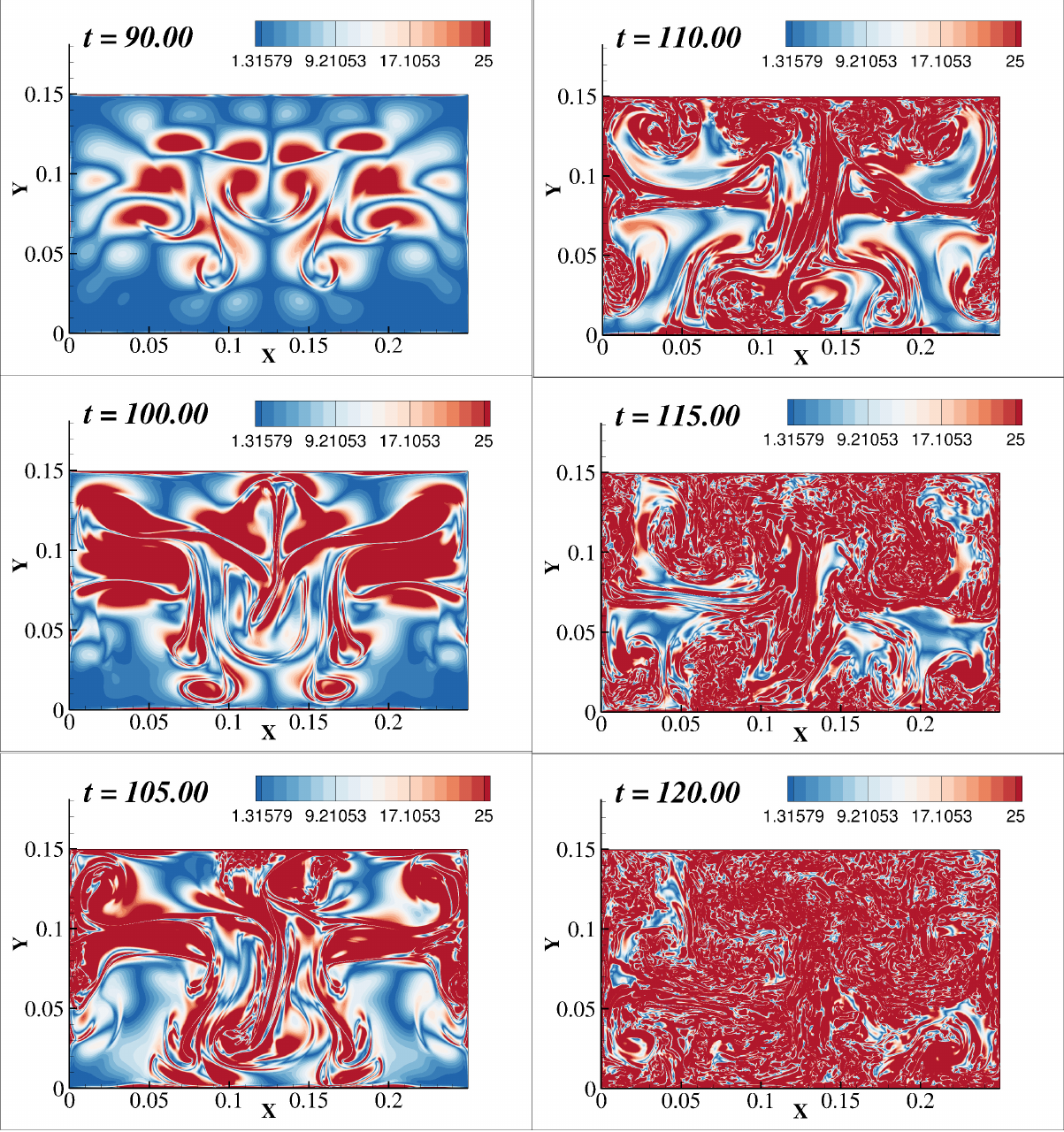}}
  \caption{Vorticity contours at different stages of case-A.}
  \label{fig6}
\end{figure}

\begin{figure}
  \centerline{\includegraphics[width=\textwidth]{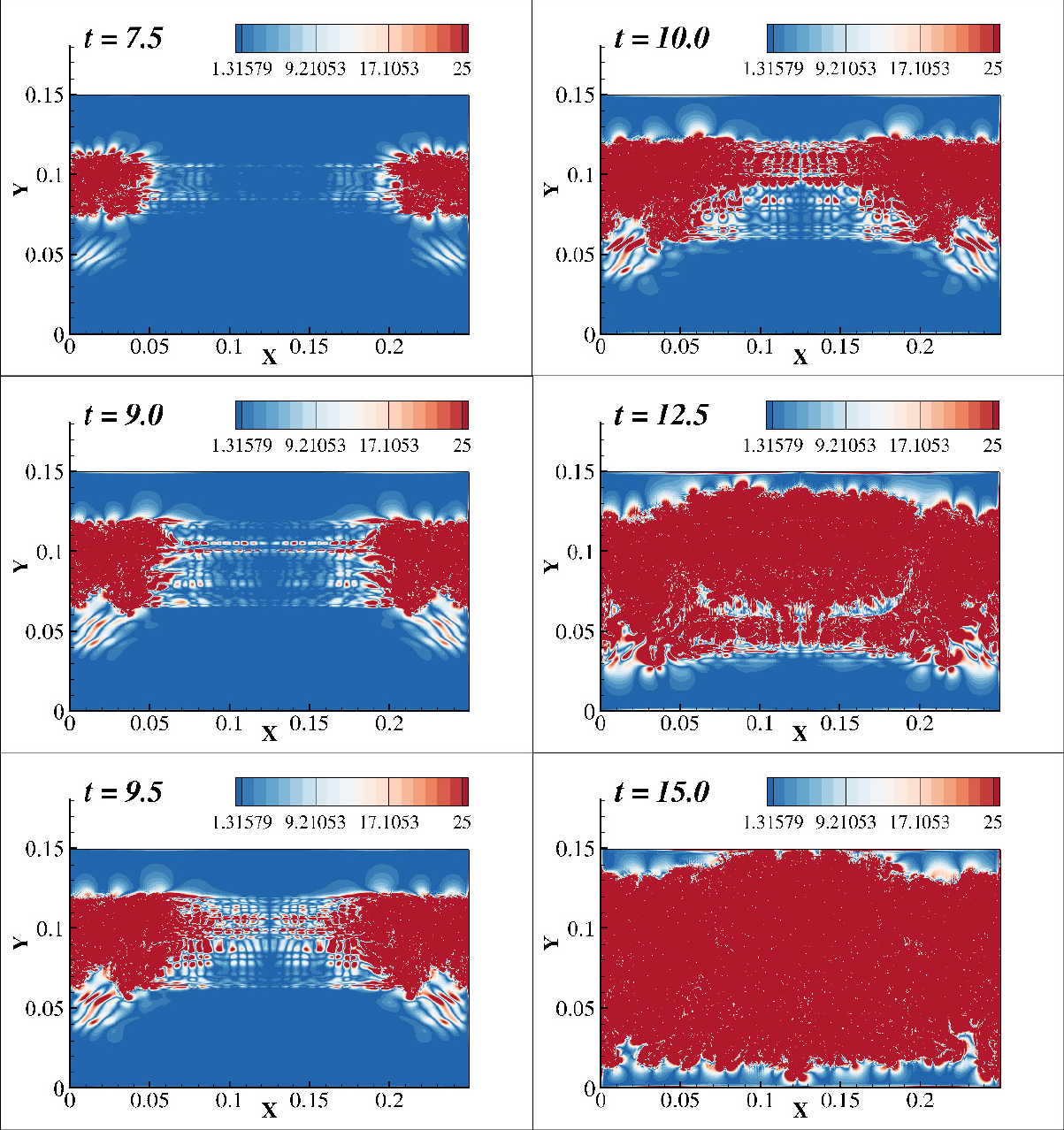}}
  \caption{Vorticity contours at different stages of case-J.}
  \label{fig7}
\end{figure}

Figures~\ref{fig6} and~\ref{fig7} present the temporal evolution of the vorticity field for the limiting weakly (Case-A) and strongly (Case-J) stratified configurations, respectively, in order to examine the distinct nonlinear pathways associated with effectively single-interface and fully coupled multi-layer RTI dynamics. While Figs. \ref{fig3}-\ref{fig5} establish the transition in the global flow topology with increasing interfacial $At$, the instantaneous contour sequences shown here provide additional insight into the evolution of coherent vortical structures, interfacial interaction, and mixing-layer merger in the two limiting regimes.

Fig. \ref{fig6} illustrates the late-stage evolution of Case-A corresponding to the weakest stratification investigated ($At_{12}=0.0083, At_{23}=0.0245$). The flow evolution closely resembles classical low-$At$ single-interface RTI characterized by gradual bubble-spike growth, coherent vortex roll-up, and delayed transition to turbulence \citep{sharp1984overview, sundaram2020effects, sengupta2020nonmodal}. At $t=90$, the flow retains large-scale coherent vortical structures driven primarily by the larger density contrast across the lower interface ($At_{23} > At_{12}$). As the instability evolves, the coherent vortices undergo progressive stretching, interaction, and fragmentation, producing elongated vortical filaments and secondary roll-up structures. Between $t=100$ and 110, strong vortex interaction leads to the formation of increasingly disordered multiscale structures occupying a substantial fraction of the computational domain. By $t=120$, the flow has transitioned into a fully developed turbulent mixing state characterized by widespread small-scale vorticity distribution.At these late stages ($t=115$-$120$), boundary-driven spatial confinement and intense non-linear interaction lead to the complete loss of the initially discrete interfacial boundaries, resulting in a fully merged, domain-wide turbulent state.

In contrast, Fig. \ref{fig7} presents the evolution of Case-J corresponding to the strongest interfacial stratification ($At_{12}=0.0555, At_{23}=0.0633$). The flow exhibits qualitatively different dynamics associated with fully coupled multi-layer RTI evolution \citep{sengupta2023multi, Suchandra_Ranjan_2023}. At $t=7.5$, intense vortical activity is established near both interfaces, with the upper interface ($y \approx 0.10$) exhibiting concentrated, compact multiscale vortex cores \citep{roberts2016effects}. This localized intensity stems from strong vertical confinement near the upper interface, which keeps the density and pressure gradients tightly misaligned ($\nabla \rho \times \nabla p$) to amplify local peak vorticity. Conversely, the lower interface ($y \approx 0.05$) experiences stronger buoyancy forces ($At_{23} > At_{12}$), driving rapid vertical fluid displacement over a broader region ($y \approx 0.025$-$0.075$). This spatial dispersion spreads out the vorticity, organizing the lower layer into elongated, slanted vortical filaments \cite{suman2022novel} rather than compact cores. Between $t=9.0$ and 10.0, strong nonlinear interaction develops between the two actively evolving mixing layers, leading to rapid broadening of the interfacial region and progressive collapse of the initially stratified intermediate layer. By $t=12.5$, the two mixing layers have merged into a single turbulent region occupying most of the domain interior, and at $t=15.0$ the flow is characterized by highly disordered small-scale vortical structures typical of fully developed buoyancy-driven turbulent mixing \cite{aslangil2023investigation, lawrie2010rayleigh}.

It is important to note that the inter-layer merging observed in these two-dimensional vorticity contours represents the visual spatial coalescence of active vortical belts. This visual structural interaction directly precedes the domain-averaged expansion of the integral mixing layer widths ($h$) evaluated quantitatively in next section. The two limiting cases overall highlight the fundamental change in instability dynamics induced by increasing interfacial stratification: weakly stratified configurations evolve through delayed coherent single-interface vortex dynamics, whereas strongly stratified configurations exhibit rapid dual-interface interaction, early mixing-layer merger, and accelerated transition toward volumetrically distributed turbulent mixing.

\subsection{Characterizing the mixing layers in multi-stratified RTI}

The mixing-layer thickness, $H$, is a fundamental metric for characterizing the temporal evolution of Rayleigh--Taylor instability (RTI), capturing the combined effects of buoyancy forcing, density stratification, and nonlinear interfacial mixing \citep{adkins2017interface}. For a single unstable interface, the quadratic growth of the mixing layer during the intermediate growth regime is classically modeled by the power law \citep{youngs1984numerical}:
\begin{equation}
H(t) = \alpha {At} g t^2,
\end{equation}
where $g$ is the gravitational acceleration, $At$ is the interfacial Atwood number, $t$ is physical time, and $\alpha$ is the dimensionless quadratic growth-rate coefficient. 

In the present numerical framework, time and spatial dimensions are normalized as $\tau = t \sqrt{g/L}$ and $h = H/L$, respectively, where $L$ is the characteristic domain length scale. Expressing the mixing-layer growth in terms of non-dimensional time $\tau$ eliminates explicit gravity dependence. To extract the growth parameters across all test cases, the non-dimensionalized mixing-layer data are fitted during the intermediate growth window using the origin-adjusted quadratic regression form:
\begin{equation}
(h - h_{\min}) = \alpha {At} (\tau - \tau_{\min})^2,
\label{mixing}
\end{equation}
where $h_{\min}$ and $\tau_{\min}$ denote the non-dimensional baseline height and non-dimensional time corresponding to the onset of the active intermediate growth phase for each respective layer. Because full self-similar turbulence is not uniformly reached across all test cases, the fitted parameters ($\alpha_1$ for the upper layer, $\alpha_2$ for the lower layer, and $\alpha_{12}$ for the combined mixing width) represent effective quadratic growth coefficients over the intermediate regime rather than late-time turbulent constants.

\begin{figure}
  \centerline{\includegraphics[width=\textwidth]{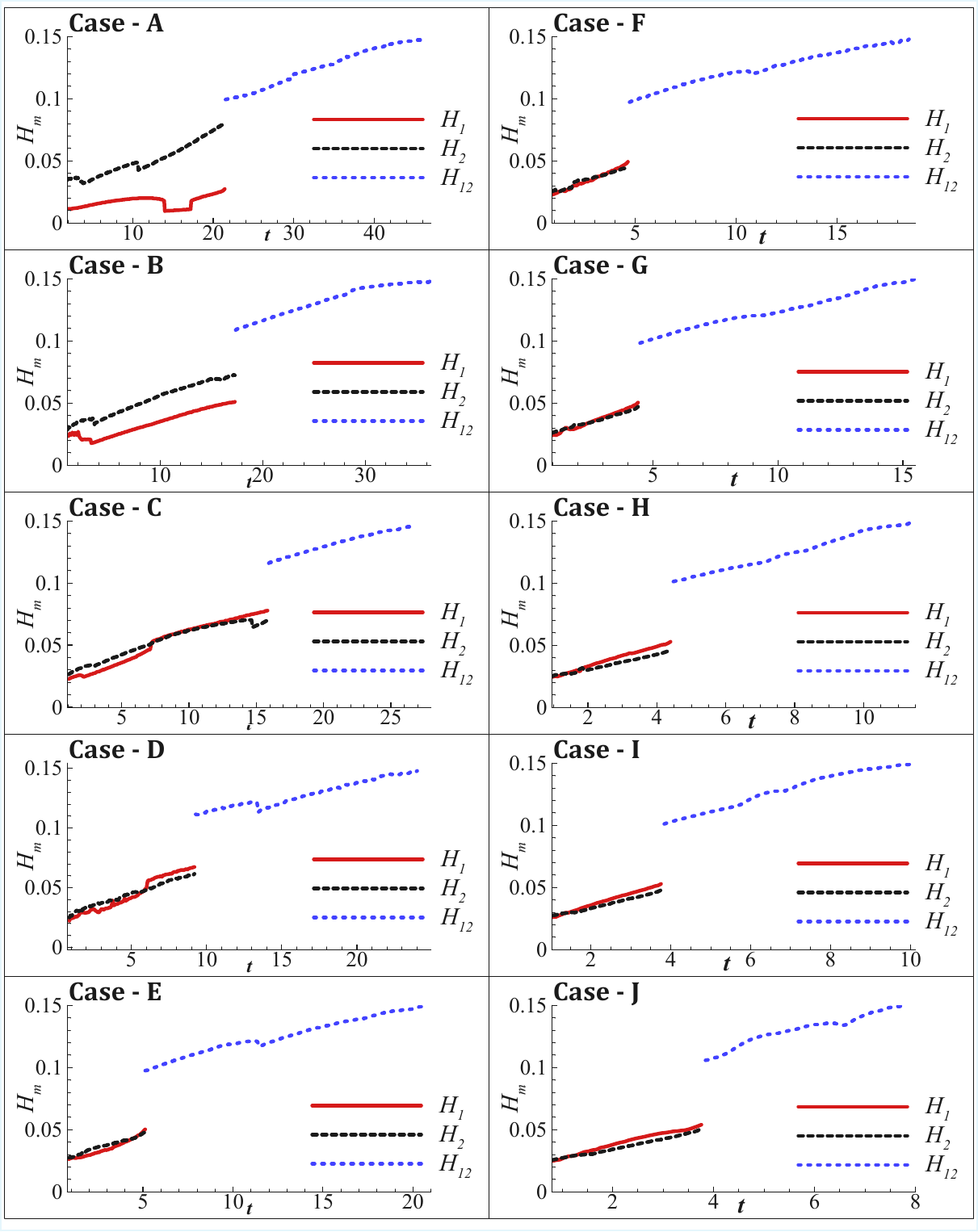}}
  \caption{Temporal evolution of the mixing layer heights for cases A-J. Red and black curves represent individual interfaces; the blue dotted curve represents the merged layer.}
  \label{fig8}
\end{figure}

Figure~\ref{fig8} presents the temporal evolution of the normalized mixing-layer heights for all investigated configurations arranged in increasing order of interfacial Atwood number from case-A to J. The individual mixing-layer heights associated with the upper and lower interfaces are denoted by $H_1$ and $H_2$, respectively, while $H_{12}$ represents the merged mixing layer formed following interaction between the two evolving instability regions. The mixing-layer thickness is evaluated by identifying the wall-normal spatial limits of the interfacial mixing zone extracted directly from the 2D vorticity contour distributions. To maintain consistency across all Atwood numbers, the boundaries of the mixing layers are identified using fixed threshold contour levels corresponding to the outermost spatial extent of the unstable interfacial structures. Prior to interface collision, $H_1$ and $H_2$ are evaluated independently across their respective interfaces, whereas $H_{12}$ is tracked continuously following the onset of interfacial interaction and subsequent merging.

At early times, the upper and lower mixing layers evolve largely independently and exhibit comparatively slow growth associated with the linear and weakly nonlinear stages of the instability. The weakly stratified configurations (Cases A-C) display limited growth of both $H_1$ and $H_2$, with the mixing regions remaining localized near their respective interfaces over extended durations. In these cases, the merged mixing layer appears only at later times and grows gradually, indicating weak interfacial coupling and limited large-scale buoyancy-driven transport. As the interfacial Atwood numbers increase (Cases D-F), the growth rates of both individual mixing layers increase substantially and the merging of mixing layers occurs progressively earlier. The enhanced density gradient strengthens baroclinic vorticity production and vertical transport, leading to stronger interaction between the two mixing regions and accelerated growth of the merged layer \citep{sharp1984overview}. Case E marks the transition toward fully coupled multi-layer dynamics, consistent with the emergence of simultaneous upper- and lower-interface activity observed previously in the vorticity fields. The strongly stratified configurations (Cases G-J) exhibit rapid development of the merged mixing region, with $H_{12}$ increasing substantially faster than the individual layer heights. In these cases, nonlinear interaction between disturbances generated at the adjacent interfaces leads to efficient redistribution of fluid across the stratified domain and rapid collapse of the initially separated intermediate layer. Although Case-J corresponds to the largest interfacial density gradient, the individual layer growth rates saturate at late times while the merged layer continues to expand steadily, suggesting increased influence of nonlinear dissipation and confinement effects \citep{cook2004mixing} during the fully developed turbulent stage. Overall, Fig.~\ref{fig8} quantitatively demonstrates the transition from weakly coupled interface-localized growth to strongly coupled multi-layer mixing with increasing interfacial stratification strength.




\begin{table}[!ht]
\caption{Growth-rate coefficients of the multi-layer RTI for cases A-J.}
\label{tab2}
\begin{ruledtabular}
\begin{tabular}{c c c c}
\textbf{Case} & $\alpha_1$ & $\alpha_2$ & $\alpha_{12}$ \\
\hline
A & 0.004856 & 0.005544 & 0.006647 \\
B & 0.012571 & 0.007856 & 0.007764 \\
C & 0.020463 & 0.008860 & 0.013903 \\
D & 0.041243 & 0.019484 & 0.007505 \\
E & 0.056026 & 0.033153 & 0.008681 \\
F & 0.075606 & 0.041676 & 0.009800 \\
G & 0.074928 & 0.043287 & 0.013688 \\
H & 0.069933 & 0.043645 & 0.028806 \\
I & 0.092678 & 0.055172 & 0.032823 \\
J & 0.075200 & 0.051606 & 0.062296 \\
\end{tabular}
\end{ruledtabular}
\end{table}

Table \ref{tab2} summarizes the growth-rate coefficients associated with the upper ($\alpha_1$), lower ($\alpha_2$), and merged ($\alpha_{12}$) mixing layers obtained from quadratic fits to the nonlinear growth regime shown in Fig.\ref{fig8}. The reported values are consistent with the range typically observed in low-to-moderate $At$ RTI \citep{youngs1984numerical, dimonte2004comparative, roberts2016effects}, while also reflecting the additional complexity introduced by interfacial coupling in the present multi-layer configuration.

The upper-interface growth coefficient, $\alpha_1$, increases progressively from Case-A to Case-I, indicating systematic enhancement of upper-layer instability growth with increasing interfacial stratification. The most pronounced increase occurs between Cases C and F, corresponding to the transition from weakly coupled to fully coupled multi-layer dynamics identified previously in the vorticity and mixing-layer evolution. Beyond Case-F, the increase in $\alpha_1$ becomes non-monotonic, with Cases G, H, and J exhibiting partial saturation relative to Case-I. This behaviour suggests that although stronger density gradients enhance buoyancy-driven growth during the early nonlinear stages, late-stage vortex interaction and turbulent dissipation increasingly moderate the individual layer growth rates at large $At$.

A similar trend is observed for the growth rate coefficient of the lower interface, $\alpha_2$, although its growth remains comparatively weaker throughout the investigated range. The coefficient increases gradually from Cases A-C, followed by a more rapid increase for Cases D-F associated with enhanced nonlinear interaction between the two mixing layers. Beyond Case G, the growth of $\alpha_2$ again becomes less pronounced, indicating progressive departure from independently evolving interface dynamics as the merged turbulent region begins to dominate the late-stage evolution.

The merged-layer coefficient, $\alpha_{12}$, exhibits the clearest signature of the transition in instability dynamics. For the weakly stratified configurations (Cases A-F), $\alpha_{12}$ remains relatively small and varies only weakly, consistent with delayed layer interaction and limited interfacial coupling. In contrast, Cases G-J exhibit a rapid increase in $\alpha_{12}$, indicating accelerated growth of the merged turbulent region following the onset of strong multi-layer interaction. The substantial increase observed for Cases H-J reflects efficient vertical redistribution of fluid and enhanced nonlinear coupling between the two actively evolving mixing layers. The exceptionally large value obtained for Case-J indicates that, although the individual interface growth rates begin to saturate at high $At$, the merged turbulent layer continues to expand efficiently owing to sustained large-scale buoyancy-driven transport and mixing.

\subsection{Asymmetric density stratification and its role in multi-layer RTI}

\begin{figure*}
    \centering
    \includegraphics[width=0.92\textwidth]{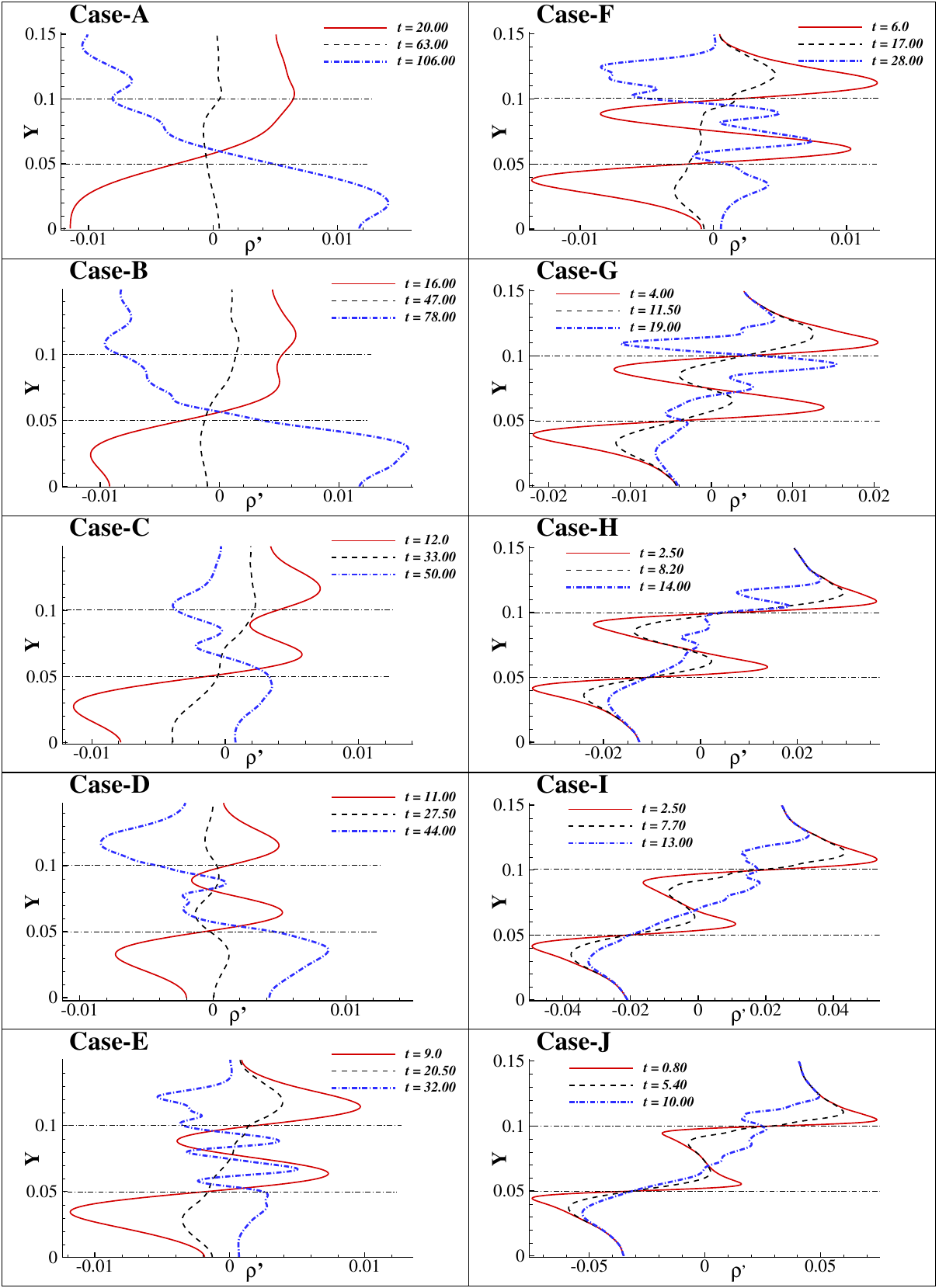}
    \caption{Variation of disturbance density $\rho'$ with the vertical coordinate $y$. Left column shows Cases A-E and right column shows Cases F-J. Profiles are shown at three stages: onset (red), log-linear growth (black), and nonlinear saturation (blue).}
    \label{fig9}
\end{figure*}

The wall-normal distribution of density fluctuations is an important diagnostic in multi-layer RTI because it directly characterizes the evolution of interfacial mixing, stratification breakdown, and vertical redistribution of fluid during the instability development. Unlike global measures such as mixing-layer height or peak vorticity, the density-fluctuation profile provides spatially resolved information regarding where mixing is localized, how rapidly the initially stratified layers lose coherence, and the extent to which the individual mixing regions interact and merge. The disturbance density, $\rho' = \rho - \bar{\rho}$, is analyzed at three representative stages of the instability development: the onset, the exponential growth regime, and start of nonlinear saturation, identified previously in Fig.~\ref{fig2}. The vertical profiles $\rho'(y)$ extracted from the domain are shown in Fig.~\ref{fig9}.

During the onset stage (red solid line), $\rho'$ remains strongly localized near the unstable interfaces located at $y = 0.05$ and 0.10, reflecting weak interfacial deformation and limited vertical transport. The weakly stratified configurations (Cases A-C) exhibit comparatively small fluctuation amplitudes confined to narrow regions surrounding the interfaces, consistent with the delayed instability growth and weak interfacial coupling observed previously in the vorticity evolution of Fig.~\ref{fig3}. In contrast, the strongly stratified cases (Cases G-J) display substantially larger disturbance amplitudes even during the onset stage, indicating enhanced baroclinic production and rapid buoyancy-driven entrainment.

As the flow evolves into the quasi-exponential growth regime (black dashed line), the fluctuation profiles broaden significantly owing to enhanced vortex-induced transport and interfacial deformation. For the weakly stratified configurations, the density fluctuations remain relatively localized and retain clear signatures of the two initially distinct interfaces, indicating that the individual mixing layers continue to evolve largely independently. Cases D and E exhibit progressively stronger overlap between the fluctuation regions associated with the upper and lower interfaces, marking the onset of appreciable multi-layer interaction.

At the onset of nonlinear saturation (blue dash-dotted line), the distinction between the weakly and strongly stratified regimes becomes particularly evident. The weakly stratified cases retain comparatively smooth and spatially separated fluctuation distributions, indicating incomplete merger of the two mixing regions and persistence of the intermediate stratified layer. In contrast, the strongly stratified configurations exhibit broad multiscale $\rho'$ distributions \citep{Suchandra_Ranjan_2023} extending across a substantial fraction of the computational domain. In Cases H-J, the profiles become highly irregular with pronounced overlap between the upper- and lower-interface fluctuation regions, indicating strong interfacial coupling, efficient vertical redistribution of fluid, and progressive destruction of the initially layered density structure \citep{aslangil2023investigation}.

The evolution of the $\rho'$ profiles in Fig. \ref{fig9} provides clear statistical evidence of the transition from weakly coupled interface-localized instability growth to fully coupled volumetrically distributed turbulent mixing with increasing interfacial stratification strength.

\subsection{Energetics, budget and scaling in multi-layer RTI}
The preceding sections established the transition from weakly coupled interface-localized instability growth to strongly coupled multi-layer turbulent mixing with increasing interfacial $At$. To further characterize the energetics and scale-dependent dynamics of the evolving flow, this section examines the temporal evolution and spatial redistribution of turbulent kinetic energy together with the associated spectral characteristics of the fully developed turbulent state. Further, the exchange between kinetic energy, potential energy, and viscous dissipation is analysed during the nonlinear stages of the flow.

\subsubsection{Turbulent kinetic energy}

The turbulent kinetic energy (TKE) is evaluated from the fluctuating velocity components as

\begin{equation}
k = \frac{1}{2} (u'^2 + v'^2 + w'^2)
\end{equation}

\noindent where $u'$, $v'$, and $w'$ denote the streamwise, wall-normal and spanwise velocity fluctuations, respectively, obtained relative to the corresponding spatially averaged mean velocities. The quantity therefore represents the kinetic energy associated with the turbulent velocity fluctuations generated during the instability evolution. The spatio-temporal evolution of TKE constitutes an important diagnostic for multi-layer RTI because it directly characterizes the generation, amplification, transport, and redistribution of turbulence arising from buoyancy-driven interfacial dynamics.

\begin{figure}[htbp]
  \centerline{\includegraphics[width=\textwidth]{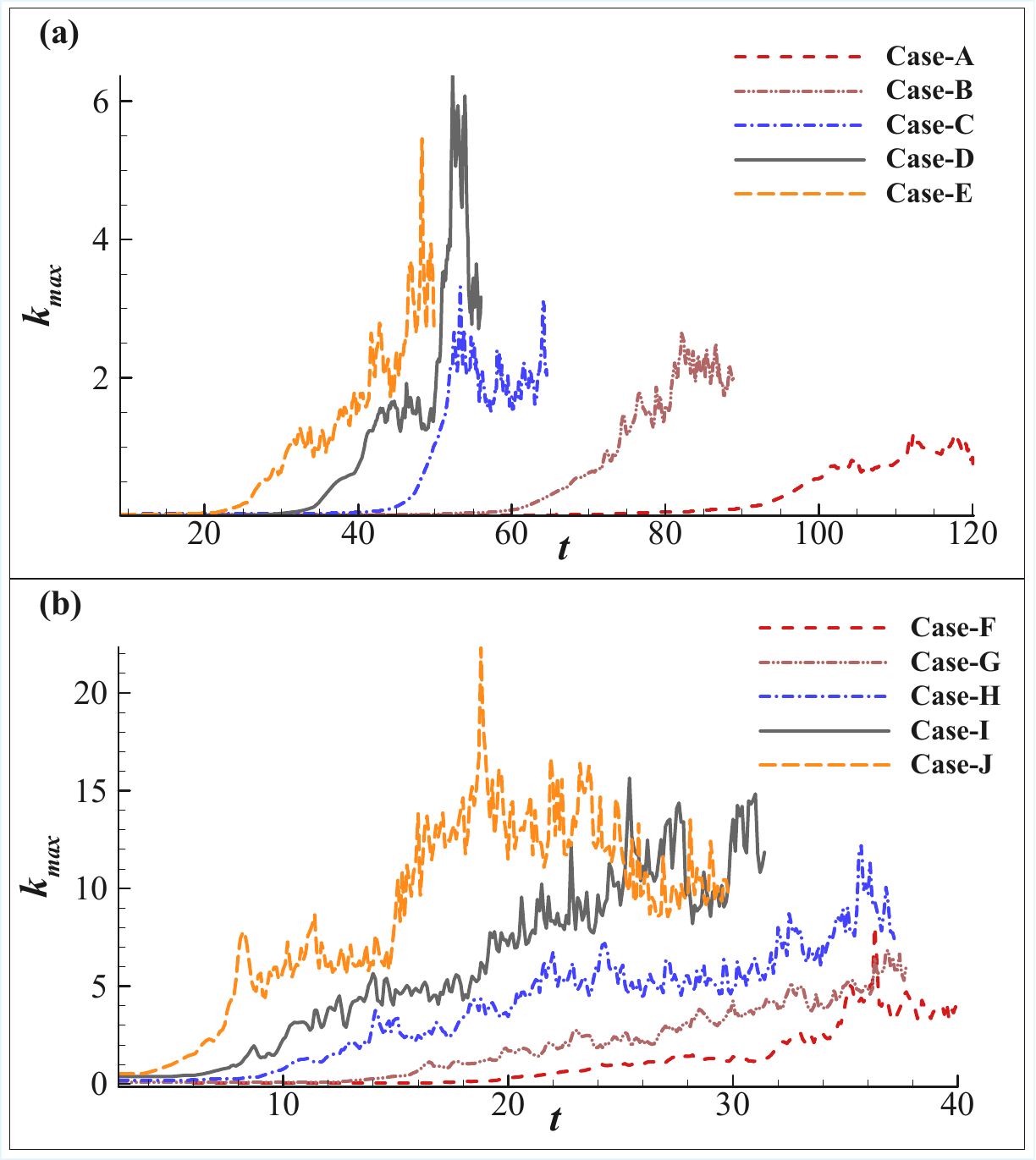}}
  \caption{Temporal evolution of the maximum turbulent kinetic energy ($k_{\max}$) for cases A-J.}
  \label{fig10}
\end{figure}

Figure~\ref{fig10} presents the temporal evolution of the maximum turbulent kinetic energy, $k_{\max}$, for all investigated configurations, with the weakly and moderately stratified cases (A-E) shown in Fig. \ref{fig10}(a) and the strongly stratified cases (F-J) shown in Fig. \ref{fig10}(b). The evolution of $k_{\max}$ provides a direct measure of the intensity and temporal development of turbulence generated during the multi-layer RTI. All configurations exhibit an initial period of weak turbulent activity followed by rapid amplification of $k_{\max}$ as the flow transitions into the nonlinear growth regime. The weakly stratified cases (A-C) display delayed growth of turbulent kinetic energy accompanied by comparatively small peak amplitudes. In particular, Case-A reaches its peak turbulent activity only at very late times ($t \approx 105\text{-}115$) with $k_{\max} \le 1$, reflecting the slow instability growth and weak interfacial coupling characteristic of effectively single-interface low-$At$ RTI \citep{sharp1984overview, youngs1984numerical}. Cases B and C exhibit progressively earlier turbulence amplification and moderately larger peak values as the interfacial density gradients increase. Cases D and E exhibit a substantial increase in turbulence intensity together with significantly earlier attainment of peak TKE levels. Case-D reaches peak values exceeding $k_{\max} \approx 6$ near $t \approx 52$, while Case-E displays strong intermittent bursts with peak values approaching $k_{\max} \approx 5$ at even earlier times. These cases correspond to the transitional regime identified in the vorticity and mixing-layer analyses of Figs. \ref{fig3}-\ref{fig5} and \ref{fig8}, where enhanced interaction between the upper and lower mixing layers leads to rapid amplification of turbulent fluctuations.

The strongly stratified cases (Cases F-J) shown in Fig. \ref{fig10}(b) exhibit markedly different behavior characterized by rapid turbulence amplification, significantly larger peak TKE values, and substantially shorter transition times. The onset of strong turbulent activity shifts progressively toward earlier times with increasing interfacial $At$, consistent with enhanced buoyancy forcing and accelerated baroclinic vorticity production \citep{cook2004mixing, joshi2025comparing}. The nonlinear regime is characterized by sustained temporal fluctuations followed by kinetic energy saturation. Cases H-J exhibit the largest peak amplitudes, with Case-J attaining $k_{\max} > 20$ at $t \approx 18.5$, indicating extremely rapid transition toward strongly nonlinear turbulent mixing. 

\begin{figure}[htbp]
  \centerline{\includegraphics[width=0.92\textwidth]{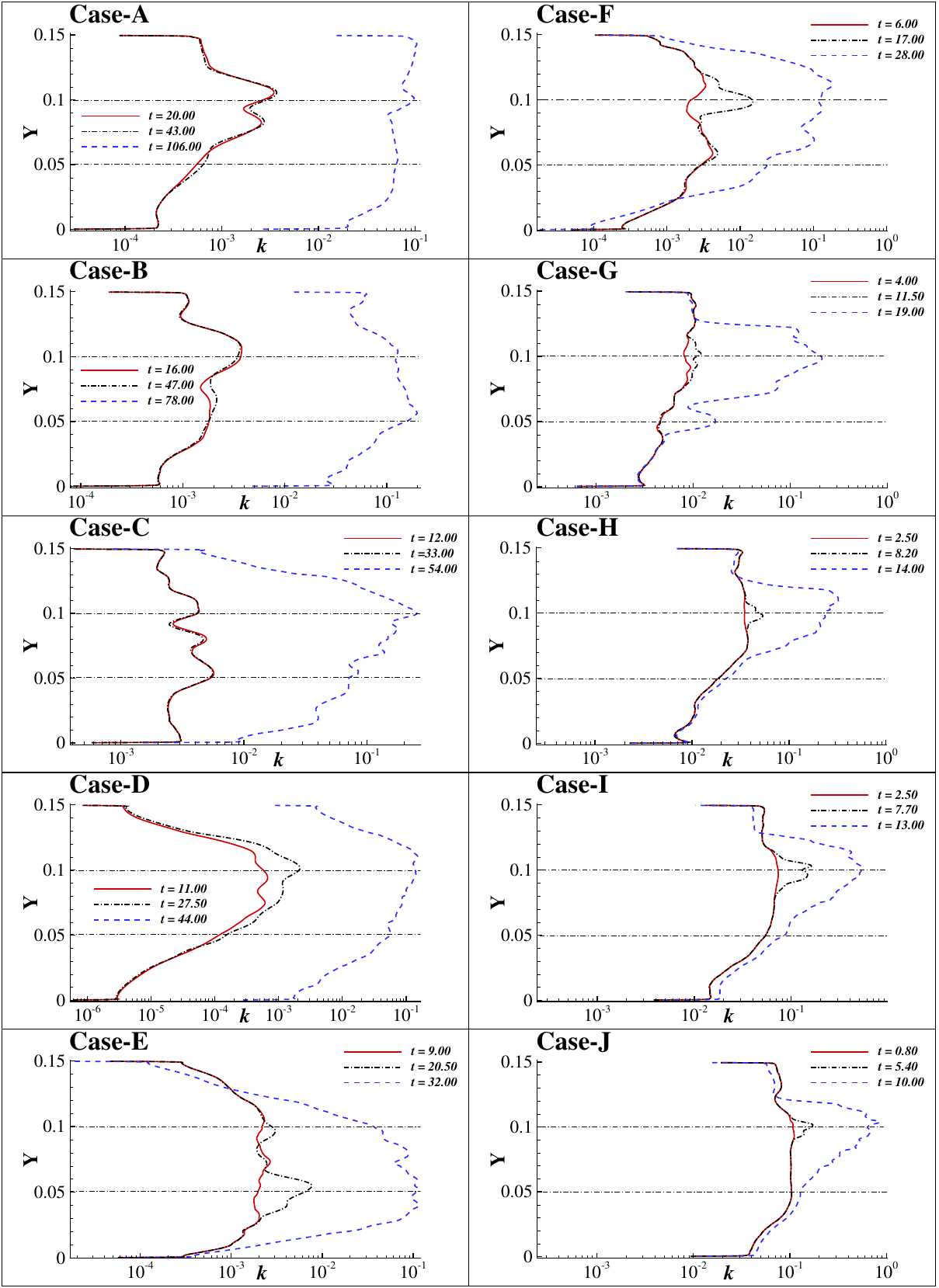}}
  \caption{Variation of turbulent kinetic energy ($k$) with vertical coordinate $y$ for cases A-J. Profiles are shown at three stages: onset (red), log-linear growth (black), and nonlinear saturation (blue). The horizontal axis is plotted on a logarithmic scale.}
  \label{fig11}
\end{figure}

The eventual saturation of $k_{\max}$ observed across all cases in Fig. \ref{fig10} reflects the physical impact of domain confinement and boundary interactions. Because the domain has a finite vertical extent and is laterally bounded, vertical fluid displacement eventually becomes constrained as the expanding mixing layer approaches the boundaries (a feature further confirmed by the wall-normal turbulent kinetic energy profiles in Fig. \ref{fig11}). Consequently, the flow does not transition into an unconfined, infinite-domain asymptotic self-similar regime at late stages. To maintain strict physical consistency, all power-law growth exponents ($\alpha$) and mixing rates extracted in this study are evaluated exclusively during the unconfined intermediate growth phase ($t < t_{\text{peak}}$), where buoyancy production dominates before boundary-enforced saturation sets in. The strongly intermittent fluctuations observed post-peak characterize a wall-bounded, domain-confined turbulent state driven by vigorous vortex collisions and multiscale spatial redistribution \citep{cabot2006reynolds, joshi2024highly}.

An intriguing feature in Fig. \ref{fig10} is the localized non-monotonicity in peak $k_{\max}$, where Case-D and Case-F attain slightly higher instantaneous maxima than Case-E and Case-G, respectively, despite the latter having higher nominal Atwood numbers. As shown in Fig. \ref{fig5}, by the non-linear saturation stage ($t = 44.00$), Case-D develops a fully interactive, vertically merged central turbulent core spanning $x \in [0.08, 0.17]$, accumulating strong velocity fluctuations across the domain interior. In contrast, for Case-E ($t = 32.00$), active vorticity generation at saturation remains predominantly localized near the domain sidewalls ($x < 0.05$ and $x > 0.20$), leaving a comparatively unmixed central core that moderates the global peak TKE. A similar mechanism governs the difference between Case-F and Case-G, where Case-F displays higher localized peak vorticity magnitude ($\text{Max} \approx 2239.76$) compared to Case-G ($\text{Max} \approx 1087.28$) due to intense vortex core concentration prior to late-stage dispersion \citep{Lawrie_Dalziel_2011}. Furthermore, the non-uniform increase in initial thermal stratification across cases (Table~\ref{tab1}), where temperature increments in $T_{\text{middle}}$ and $T_{\text{lower}}$ become significantly larger in cases H-J compared to A-G, alters the relative temporal evolution of the two density interfaces and influences the precise timing and spatial localization of peak kinetic energy production.


Figure~\ref{fig11} presents the wall-normal distributions of $k$ for all investigated cases at the three representative stages of instability evolution corresponding to the onset, quasi-exponential growth, and nonlinear saturation regimes identified previously in Fig. \ref{fig2}. The weakly and moderately stratified configurations (Cases A-E) are shown in the left column, while the strongly stratified cases (Cases F-J) are shown in the right column. The profiles provide a quantitative measure of the spatial redistribution of turbulent activity and the progressive transition from interface-localized instability growth to volumetrically distributed turbulent mixing. The original interface locations are indicated by dotted lines, approximately at $y \approx 0.05$ and $\approx 0.10$.

During the onset stage (red solid lines), the turbulent kinetic energy remains strongly localized near the unstable density interfaces owing to the initially confined nature of the baroclinic vorticity generation. Crucially, during both onset and quasi-exponential growth, the TKE profiles decay by several orders of magnitude toward the domain boundaries ($y = 0$ and $y = 0.15$), confirming that early-to-intermediate growth occurs in a physically unconfined state, free of wall influence. The weakly stratified cases (A-C) exhibit comparatively small TKE amplitudes, with peak values remaining below $10^{-2}$ and localized predominantly near the lower interface. The corresponding wall-normal profiles are relatively narrow, indicating limited vertical transport and weak interfacial interaction. In contrast, the strongly stratified configurations (Cases H-J) display substantially larger onset-stage TKE amplitudes together with broader distributions extending across both interfaces, reflecting enhanced buoyancy forcing and accelerated turbulence production \citep{dimonte2004comparative, sengupta2022thermally}.

As the instability evolves into the quasi-exponential growth regime (black dash-dotted lines), the TKE amplitudes increase for all cases owing to rapid vortex amplification and interfacial entrainment. The weakly stratified cases continue to exhibit localized TKE peaks concentrated near the original interface locations, indicating that the two mixing layers remain only weakly coupled. Cases D and E exhibit appreciable broadening of the TKE distributions together with stronger overlap between the upper- and lower-interface peaks, consistent with the onset of multi-layer interaction identified previously in Fig. \ref{fig4}. The strongly stratified configurations exhibit markedly larger TKE amplitudes during this stage, with the peak values shifting progressively toward the upper half of the domain. This behavior reflects enhanced vertical redistribution of turbulent energy and stronger nonlinear interaction between the two actively evolving mixing layers, while remaining well-separated from boundary interactions.

At the onset of nonlinear saturation (blue dashed lines), the distinction between the weakly and strongly stratified regimes becomes particularly evident, as wall-normal spreading causes the mixing region to encompass the finite domain bounds. Cases A-C retain comparatively moderate TKE amplitudes with relatively broad but still partially localized distributions, indicating incomplete merger of the two mixing regions. In contrast, Cases F-J exhibit large-amplitude TKE distributions extending over a substantial fraction of the computational domain, characteristic of fully developed volumetrically distributed turbulent mixing. This wall-normal expansion directly demonstrates how domain confinement sets in at late stages, arresting unconstrained vertical growth and driving the TKE saturation observed in Fig. \ref{fig10}. The peak TKE amplitudes increase systematically with interfacial $At$, with Cases I and J exhibiting peak values approaching $O(10^{0})$. In these strongly stratified configurations, the dominant TKE peaks are consistently located near or above the upper interface ($y \approx 0.10$), despite the larger density gradient across the lower layer. This upward shift of the maxima indicates substantial nonlinear redistribution of turbulent kinetic energy during the coupled multi-layer evolution \citep{guo2017rayleigh}, further supported by the tight vertical spatial confinement near $y \approx 0.10$ that maintains sharp baroclinic torque ($\nabla \rho \times \nabla p$).

Overall, the wall-normal TKE distributions demonstrate that increasing interfacial density gradient not only enhances the magnitude of turbulence generation but also fundamentally alters its spatial organization. The weakly stratified cases remain dominated by localized interface-confined turbulent activity, whereas the strongly stratified configurations evolve toward broad volumetric turbulence characterized by strong interfacial coupling and efficient vertical redistribution of turbulent energy, ultimately transitioning from unconfined early-stage growth into a domain-constrained late-stage turbulent mixing regime.

\subsubsection{Potential energy release}

In RTI, the initially stratified configuration represents a gravitationally unstable state in which denser fluid overlies lighter fluid. Such a configuration stores gravitational potential energy that becomes available for conversion once the interfaces are perturbed \citep{zhou2017a_rayleigh}. As the instability develops, buoyancy forces drive the heavier fluid downward and the lighter fluid upward, resulting in progressive release of the stored potential energy through interfacial deformation and fluid motion. The released potential energy is first converted into kinetic energy through buoyancy-driven acceleration of the flow. During the linear and early nonlinear stages, this energy conversion is associated primarily with baroclinic vorticity generation arising from the misalignment of pressure and density gradients at the unstable interfaces \citep{chandrasekhar1961hydrodynamic}. The resulting vortical structures amplify rapidly and generate increasingly strong velocity fluctuations, leading to growth of $k$ within the evolving mixing layers. As the instability transitions into the nonlinear regime, coherent vortices interact, stretch, and fragment into progressively smaller scales. This process transfers kinetic energy from large buoyancy-driven structures to smaller turbulent eddies through the turbulent cascade mechanism \citep{kolmogorov1991dissipation}. At sufficiently small scales, viscous effects become dominant and the kinetic energy is irreversibly dissipated into internal energy through viscous dissipation. Consequently, the gravitational potential energy released by the unstable density stratification is partitioned between the production of turbulent kinetic energy and viscous dissipation.

In multi-layer RTI, this energy conversion process becomes more complex owing to the simultaneous evolution and interaction of multiple unstable interfaces. The relative partitioning of released potential energy into turbulent transport and viscous dissipation therefore provides important insight into the efficiency and dynamics of multi-layer turbulent mixing. To quantify this an energetics framework similar to that employed by Suchandra and Ranjan~\cite{Suchandra_Ranjan_2023} is adopted, while explicitly retaining anisotropy by accounting for all resolved velocity components and the mean density field.

The gravitational potential energy per unit width at a given streamwise location is defined using the mean density as
\begin{equation}
PE(x) = \int \bar{\rho} \, g \, y \, dy,
\end{equation}
where \( \bar{\rho} \) denotes the mean density, \( g \) is the gravitational acceleration, and \( y \) is the vertical coordinate.

The total turbulent kinetic energy per unit width is computed as
\begin{equation}
KE(x) = \int k \, dy,
\end{equation}

Assuming a step change in density at the initial interface, the reference potential energy \( PE_0 \) and turbulent kinetic energy \( KE_0 \) are evaluated from the initial condition. The total potential energy released is then defined as
\begin{equation}
\Delta PE(x) = PE_0 - PE(x),
\end{equation}
while the increase in turbulent kinetic energy is given by
\begin{equation}
\Delta KE(x) = KE(x) - KE_0.
\end{equation}
The dissipated energy is obtained as the remainder,
\begin{equation}
\Delta DE(x) = \Delta PE(x) - \Delta KE(x).
\end{equation}

\begin{table}[!ht]
\caption{Energy fraction ratios at characteristic stages of the instability.}
\label{tab3}
\begin{ruledtabular}
\begin{tabular}{c c c c c c c}
\textbf{Case} & \multicolumn{2}{c}{\textbf{Onset stage}} & \multicolumn{2}{c}{\textbf{Exponential growth}} & \multicolumn{2}{c}{\textbf{Nonlinear saturation}} \\
& $\Delta \mathrm{KE}/\Delta \mathrm{PE}$ & $\Delta \mathrm{DE}/\Delta \mathrm{PE}$ & $\Delta \mathrm{KE}/\Delta \mathrm{PE}$ & $\Delta \mathrm{DE}/\Delta \mathrm{PE}$ & $\Delta \mathrm{KE}/\Delta \mathrm{PE}$ & $\Delta \mathrm{DE}/\Delta \mathrm{PE}$ \\
\hline
A & 0.0140 & 0.9860 & 0.0141 & 0.9859 & 0.1007 & 0.8993 \\
B & 0.0139 & 0.9861 & 0.0140 & 0.9860 & 0.1062 & 0.8938 \\
C & 0.0151 & 0.9849 & 0.0146 & 0.9854 & 0.2794 & 0.7206 \\
D & 0.0007 & 0.9973 & 0.0017 & 0.9983 & 0.1646 & 0.8353 \\
E & 0.0137 & 0.9863 & 0.0162 & 0.9838 & 0.1669 & 0.8331 \\
F & 0.0262 & 0.9738 & 0.0296 & 0.9704 & 0.1812 & 0.8188 \\
G & 0.0749 & 0.9251 & 0.0760 & 0.9240 & 0.1829 & 0.8181 \\
H & 0.1490 & 0.8510 & 0.1506 & 0.8494 & 0.2645 & 0.7355 \\
I & 0.0936 & 0.9064 & 0.1003 & 0.8997 & 0.2479 & 0.7521 \\
J & 0.2668 & 0.7332 & 0.2681 & 0.7319 & 0.4940 & 0.5060 \\
\end{tabular}
\end{ruledtabular}
\end{table}

Table \ref{tab3} summarizes the normalized energy partitioning during the onset, quasi-exponential growth, and nonlinear saturation stages of the multi-layer RTI evolution. The quantities \( \Delta KE/\Delta PE \) and \( \Delta DE/\Delta PE \) represent the fractions of released gravitational potential energy converted into turbulent kinetic energy and dissipated energy, respectively. The reported values therefore quantify the efficiency with which the unstable density stratification transfers buoyancy-driven potential energy into turbulent motion across the different interfacial stratification regimes.

During the onset stage, the weakly stratified configurations (Cases A-E) exhibit extremely small values of \( \Delta KE/\Delta PE \), typically of order $10^{-2}$, indicating that only a small fraction of the released potential energy is converted into turbulent motion during the initial stages of instability development. Consequently, the dominant fraction of the released energy appears in the dissipative component, with \( \Delta DE/\Delta PE \) $\approx 0.98-0.99$. This behaviour is consistent with the early-stage flow structure observed previously in Figs. \ref{fig3} and \ref{fig11}, where the instability remains localized near the interfaces and turbulent activity is comparatively weak. This is also consistent with the findings of ~\cite{Suchandra_Ranjan_2023}, where low conversion efficiencies of potential energy into turbulent kinetic energy was reported for multi-layer RTI. Case D exhibits particularly small energy-conversion efficiency during the onset and exponential stages, suggesting delayed development of sustained turbulent fluctuations despite increased interfacial forcing. As the interfacial $At$ increases, the fraction of released potential energy converted into $k$ increases systematically. Cases F-J exhibit substantially larger values of \( \Delta KE/\Delta PE \) even during the onset stage, reflecting enhanced buoyancy-driven acceleration and more efficient production of turbulent fluctuations. In particular, Cases H-J display a marked increase in energy-conversion efficiency, with Case J converting nearly 27\% of the released potential energy directly into $k$ during the onset and exponential-growth regimes. The corresponding reduction in \( \Delta DE/\Delta PE \) indicates that progressively less energy is dissipated directly as the stratification strength increases.

The transition from the quasi-exponential growth stage to nonlinear saturation is accompanied by a substantial increase in \( \Delta KE/\Delta PE \) for all configurations. This increase reflects the rapid amplification of turbulent fluctuations associated with strong vortex interaction, mixing-layer merger, and redistribution of energy across multiple spatial scales \citep{zhou2017b_rayleigh}. The weakly stratified configurations continue to exhibit relatively modest turbulent conversion efficiencies during saturation, with dissipation remaining the dominant energy pathway. In contrast, the strongly stratified cases exhibit significantly enhanced turbulent energy retention. Cases H-J in particular display large increases in \( \Delta KE/\Delta PE \), indicating efficient conversion of gravitational potential energy into volumetrically distributed turbulent motion during the fully nonlinear stage. Case-J exhibits the largest turbulent conversion efficiency throughout the instability evolution, attaining \( \Delta KE/\Delta PE \) $\approx 0.49$ during nonlinear saturation. This indicates that nearly half of the released gravitational potential energy is retained within the turbulent flow field, while the remaining fraction is dissipated. Such behaviour is consistent with the strong multi-layer coupling, rapid mixing-layer merger, and broad turbulent kinetic energy distributions observed previously for the highest interfacial $At$. 

Overall, the results in Table \ref{tab3} demonstrate that increasing interfacial stratification fundamentally alters the energetics of multi-layer RTI by progressively shifting the energy balance from dissipation-dominated evolution toward more efficient buoyancy-driven turbulent energy production.

\subsubsection{Scaling in multi-layer RTI with low and high stratification levels}

\begin{figure}
  \centerline{\includegraphics[width=0.92\textwidth]{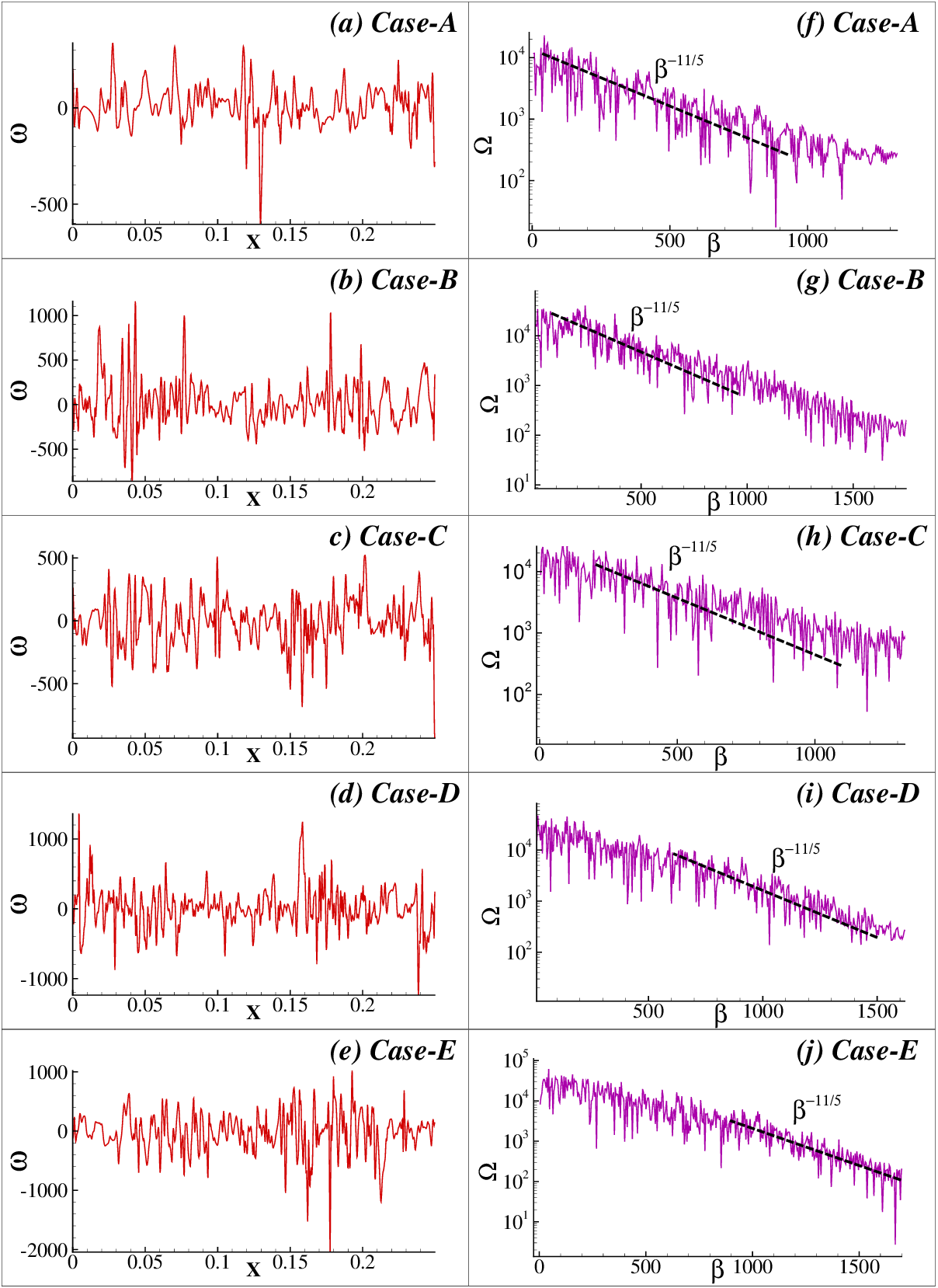}}
  \caption{Streamwise distribution of vorticity ($\omega$) and its FFT for cases A-E at the final simulation time, shown in frames (a)–(e) and (f)–(j), respectively. The profile is extracted at half-height of the domain.}
  \label{fig12}
\end{figure}

\begin{figure}
  \centerline{\includegraphics[width=0.92\textwidth]{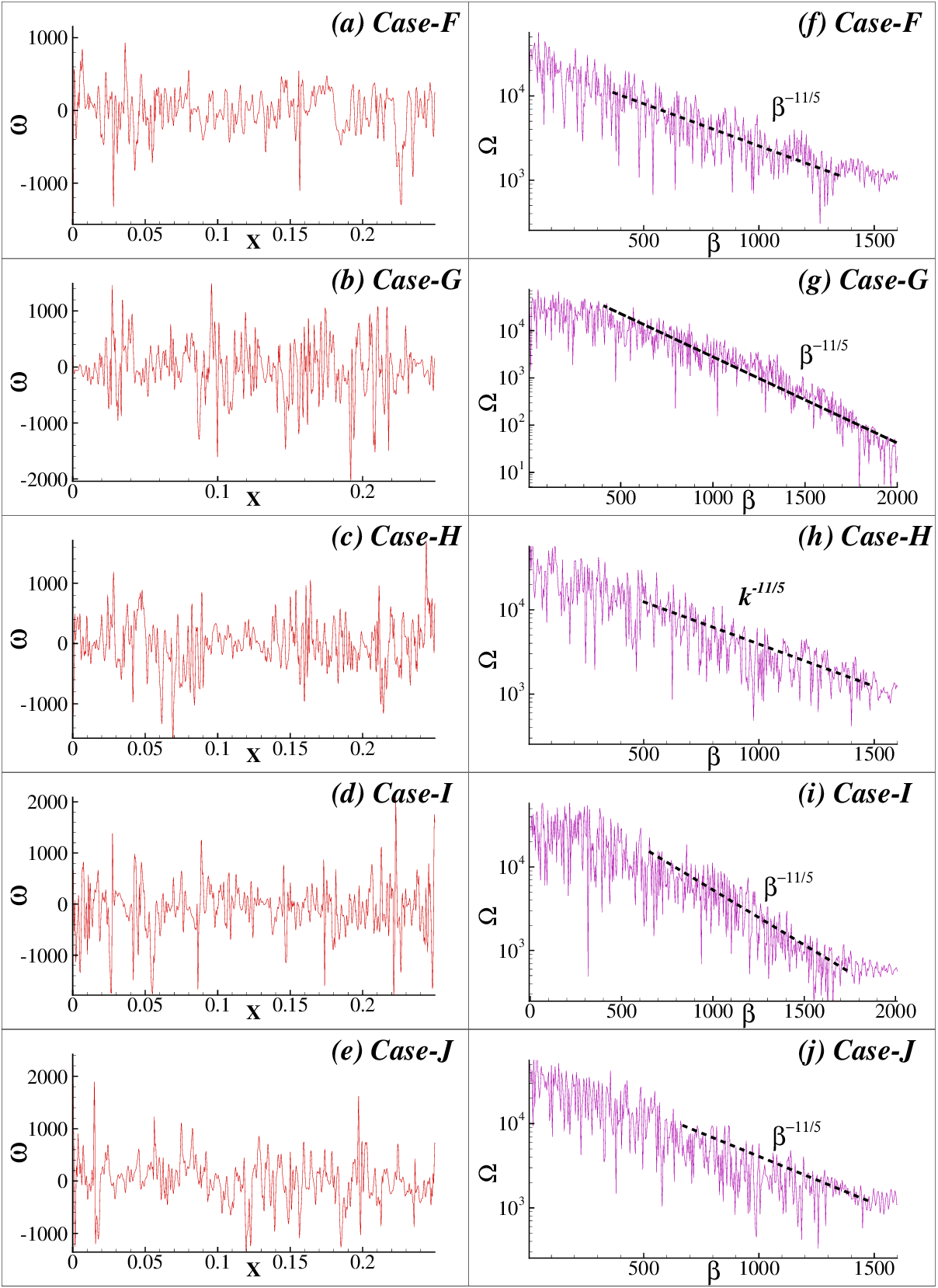}}
  \caption{Streamwise distribution of vorticity ($\omega$) and its FFT for cases F-J at the final simulation time, shown in frames (a)–(e) and (f)–(j), respectively. The profile is extracted at half-height of the domain.}
  \label{fig13}
\end{figure}

Figures \ref{fig12} and \ref{fig13} present the streamwise distributions of the vorticity field, $\omega (x)$, extracted along the horizontal centerline of the computational domain (left) together with the corresponding one-dimensional Fourier spectra (right) at the final simulation time. Cases A-E are shown in Fig.\ref{fig12}, while the strongly stratified configurations (Cases F-J) are presented in Fig.\ref{fig13}. The associated Fourier amplitudes are plotted as functions of the wavenumber, $\beta$, in order to characterize the scale-dependent structure of the fully developed multi-layer RTI flow field and to examine the emergence of buoyancy-dominated turbulent scaling.

The instantaneous vorticity distributions exhibit progressively increasing multiscale complexity with increasing interfacial $At$. The weakly stratified cases (A-C) display comparatively smoother large-scale vortical variations with limited fine-scale intermittency, consistent with the delayed and weakly coupled turbulent evolution observed previously. In contrast, the strongly stratified configurations (Cases H-J) exhibit highly irregular broadband vorticity distributions characterized by intense small-scale fluctuations and intermittent localized peaks, indicating enhanced nonlinear vortex interaction and fully developed turbulent mixing. The corresponding Fourier spectra exhibit an approximate $\Omega(\beta) \sim \beta^{-11/5}$ scaling over an intermediate range of wavenumbers for all investigated configurations. This scaling is characteristic of buoyancy-dominated turbulent flows and is consistent with energy transfer reported for stratified turbulence \citep{brizzolara2021transition, sengupta2023effects}. The emergence of the -11/5 scaling indicates that the late-stage dynamics are governed predominantly by buoyancy-driven multiscale turbulent interactions rather than isolated interfacial instability mechanisms.

For the weakly stratified configurations (Cases A-C) in Fig.\ref{fig12}, the -11/5 behaviour is confined primarily to moderate wavenumbers, approximately $\beta \sim 300-900$, beyond which the spectra exhibit stronger deviations associated with limited inertial-range development and comparatively weak small-scale turbulence generation. As the interfacial density gradients increases (Cases D-G), the scaling region broadens substantially and extends toward larger wavenumbers owing to enhanced vortex stretching, interfacial interaction, and turbulent cascade development.

The strongly stratified configurations (Cases H-J) in Fig.\ref{fig13} exhibit the most extended and well-defined scaling ranges, with the -11/5 behaviour persisting over nearly the entire intermediate-to-high-wavenumber range ($\beta \sim 300-1500$). This behaviour reflects the establishment of fully developed buoyancy-driven turbulence characterized by efficient transfer of energy across a broad spectrum of spatial scales. The stronger interfacial density gradients not only enhance the intensity of the turbulent mixing but also promote the development of broader scale-invariant spectral behaviour associated with mature stratified turbulence \citep{mikaelian1982rayleigh}.

\section{Summary and conclusions}
Direct numerical simulations have been performed to investigate the influence of interfacial $At$ on the evolution of thermally stratified multi-layer RTI. The results demonstrate that increasing interfacial density gradients fundamentally alters not only the instability growth rate, but also the structure, energetics, and scale-dependent organization of the resulting turbulent flow. Across all investigated configurations, the instability evolution proceeds through distinct onset, quasi-exponential growth, and nonlinear saturation stages; however, the nature of the flow dynamics within these regimes changes significantly with increasing stratification strength.

The simulations reveal the existence of two qualitatively distinct multi-layer instability regimes. The weakly stratified configurations evolve through dynamics that remain effectively single-interface in nature despite the presence of two unstable density interfaces. In these cases, the lower interface dominates the instability evolution, the upper interface remains comparatively passive, and the resulting turbulent activity remains localized near the primary mixing layer. In contrast, sufficiently strong interfacial stratification activates both interfaces simultaneously, producing strongly coupled multi-layer dynamics characterized by rapid interfacial interaction, accelerated mixing-layer merger, and volumetrically distributed turbulence. The transition between these regimes occurs progressively through the intermediate stratification cases, where appreciable interaction between the two evolving mixing layers first emerges.

The mixing-layer evolution and growth-rate analysis demonstrate that the merged turbulent layer becomes increasingly dominant with increasing interfacial $At$. While the individual mixing-layer growth rates exhibit partial saturation at the highest stratification levels, the merged layer continues to expand efficiently owing to enhanced nonlinear coupling and large-scale buoyancy-driven transport. The wall-normal distributions of density fluctuations and turbulent kinetic energy further show that strong stratification leads to rapid destruction of the initially layered density structure and substantial redistribution of turbulent activity across the computational domain. 

An important physical observation in this multi-layer system is that the strongest late-stage turbulent activity consistently develops near the upper interface ($y \approx 0.10$), despite the nominal Atwood number being larger across the lower interface ($At_{23} > At_{12}$). This behavior is governed by a fundamental difference in spatial confinement versus dispersion between the two layers. While the stronger buoyancy force at the lower interface ($y \approx 0.05$) induces rapid early growth, it also causes the vortical structures to undergo broad vertical spatial dispersion over $y \approx 0.025$-$0.07$, which dilutes local peak vorticity. In contrast, the upper interface experiences strong vertical spatial confinement, maintaining sharp local density and pressure misalignments ($\nabla \rho \times \nabla p$). This localized gradient sharpness dramatically amplifies baroclinic torque within compact zones, making the upper interface the primary driver of rapid non-linear vortex roll-up, localized breakdown, and subsequent domain-wide turbulent merging.

The energetics analysis demonstrates that increasing interfacial stratification substantially enhances the efficiency of conversion of gravitational potential energy into turbulent kinetic energy. Weakly stratified configurations remain strongly dissipation-dominated throughout the instability evolution, whereas the strongly stratified cases retain a significantly larger fraction of the released potential energy within the turbulent flow field. This transition in the energy partitioning behaviour reflects the increasing importance of coupled buoyancy-driven turbulence and multiscale vortex interaction at large interfacial density gradients. The late-stage spectral analysis reveals that the fully developed multi-layer RTI exhibits an approximate $-11/5$ scaling over an intermediate range of wavenumbers, consistent with buoyancy-dominated turbulent transport. Increasing interfacial $At$ broadens the extent of this scaling range owing to enhanced nonlinear interaction and stronger multiscale turbulence generation. The emergence of similar buoyancy-controlled spectral behaviour across all investigated configurations suggests that, despite substantial differences in the early-stage instability dynamics, the late-stage evolution approaches a common buoyancy-driven turbulent state.

The present study highlights the important role of interfacial coupling in governing the transition from localized instability growth to volumetrically distributed turbulent mixing in multi-layer RTI. The results also demonstrate that the nonlinear interaction between adjacent unstable interfaces can substantially modify the spatial organization and energetics of the flow relative to canonical single-interface RTI. Future investigations may focus on the influence of compressibility, miscibility effects, and three-dimensional turbulence dynamics on the coupled instability evolution. Extension of the present framework to shock-driven multi-layer Richtmyer-Meshkov instabilities and continuously stratified buoyancy-driven turbulence may also provide further insight into the dynamics of multiscale mixing processes relevant to atmospheric flows, astrophysical plasmas, inertial confinement fusion, and high-energy-density fluid systems.

\begin{acknowledgments}
The authors would like to acknowledge the use of high-performance computing facilities: ARYABHATA  and Chandrasekhar (0.5 million core-hours each) at IIT (ISM) Dhanbad, and the French high performance computing cluster Nautilus at Ecole Centrale de Nantes (1.36 million core-hours) for all the computations reported here. This research was funded by Department of Science and Technology (DST), Government of India, and the Indo-French Centre for the Promotion of Advanced Research (CEFIPRA) under Grant 69T05-1. 
\end{acknowledgments}

\section*{Declaration of Interests}
The authors report no conflict of interest.

\section*{Data Availability}

The data will be made available upon reasonable request.


\bibliography{multi_layer_rti}


\end{document}